\documentclass[english,12pt]{article}
\usepackage{epsfig,amssymb,amsmath}
\usepackage{array}
\usepackage{multirow}
\usepackage{graphicx}
\usepackage{natbib}
\usepackage{verbatim}
\usepackage{babel}
\usepackage{amsthm}
\usepackage{setspace}
\newcommand{\beq}{\begin{equation}}
\newcommand{\eeq}{\end{equation}}
\newcommand{\kms}{\,km~s$^{-1}$ }

\newcommand{\hmpc}{\,$h^{-1}$Mpc }

\newcommand{\avg}[1]{\ensuremath{\langle{#1}\rangle}}
\newcommand{\hia}{H\,{\sc i}}
\newcommand{\hi}{H\,{\sc i}~}
\newcommand{\kmsa}{km ${\rm s^{-1}}$}

\bibpunct[; ]{(}{)}{,}{a}{}{;}
\def\Title#1{\begin{center} {\Large {\bf #1} } \end{center}}
\begin{document}
\hbox to\hsize{\null}

\vfill
\begin{center}

\begin{LARGE}
{\bf
Proceedings of the 2011 \\[1ex]
New York Workshop on\\[1ex]
Computer, Earth and Space Science}
\end{LARGE}

\vfill

\vfill\vfill

\begin{Large}
February 2011\\[2ex]
Goddard Institute for Space Studies\\
\bigskip
http://www.giss.nasa.gov/meetings/cess2011\\

\vfill\vfill \vfill

Editors\\
M.J. Way and C. Naud\\
\end{Large}
\vfill\vfill

Sponsored by the Goddard Institute for Space Studies
\end{center}

\newpage
\noindent {\bf \Large Contents}\par\bigskip
\noindent {\bf Foreword}\par
{\it Michael Way \& Catherine Naud } \hfill 4\par
\bigskip
\noindent {\bf Introduction}\par
{\it Michael Way} \hfill 5\par
\bigskip
\noindent {\bf On a new approach for estimating threshold crossing times\newline
with an application to global warming}\par
{\it Victor H. de la Pe\~{n}a} \hfill 8\par
\bigskip
\noindent {\bf Cosmology through the large-scale structure of the Universe}\par
{\it Eyal Kazin} \hfill 13\par
\bigskip
\noindent {\bf On the Shoulders of Gauss, Bessel, and Poisson: Links,\newline Chunks,
Spheres, and Conditional Models}\par
{\it William Heavlin} \hfill 21\par
\bigskip
\noindent {\bf Mining Citizen Science Data: Machine Learning Challenges}\par
{\it Kirk Borne} \hfill 24\par
\bigskip
\noindent {\bf Tracking Climate Models}\par
{\it Claire Monteleoni} \hfill 28\par
\bigskip
\noindent {\bf Spectral Analysis Methods for Complex Source Mixtures}\par
{\it Kevin Knuth} \hfill 31\par
\bigskip
\noindent {\bf Beyond Objects: Using Machines to Understand the Diffuse\newline
Universe}\par
{\it J.E.G. Peek} \hfill 35\par
\bigskip
\noindent {\bf Viewpoints: A high-performance high-dimensional exploratory\newline
data analysis tool}\par
{\it Michael J. Way} \hfill 44\par
\bigskip
\noindent {\bf Clustering Approach for Partitioning Directional Data in\newline
Earth and Space Sciences}\par
{\it Christian Klose} \hfill 46\par
\bigskip
\noindent {\bf Planetary Detection: The Kepler Mission}\par
{\it Jon Jenkins} \hfill 52\par
\bigskip
\noindent {\bf Understanding the possible influence of the solar activity\newline
on the terrestrial climate: a times series analysis approach}\par
{\it Elizabeth Mart\'{i}nez-G\'{o}mez} \hfill 54\par
\bigskip
\noindent {\bf Optimal Scheduling of Exoplanet Observations Using Bayesian\newline
Adaptive Exploration}\par
{\it Thomas J. Loredo} \hfill 61\par
\bigskip
\noindent {\bf Beyond Photometric Redshifts using Bayesian Inference}\par
{\it Tam\'{a}s Budav\'{a}ri} \hfill 65\par
\bigskip
\noindent {\bf Long-Range Climate Forecasts Using Data Clustering and\newline Information Theory}\par
{\it Dimitris Giannakis} \hfill 68\par
\bigskip
\noindent {\bf Comparison of Information-Theoretic Methods to estimate\newline
the information flow in a dynamical system}\par
{\it Deniz Gencaga} \hfill 72\par
\bigskip
\noindent {\bf Reconstructing the Galactic halo’s accretion history: A finite\newline
mixture model approach}\par
{\it Duane Lee \& Will Jessop} \hfill 77\par
\bigskip
\noindent {\bf Program} \hfill 78\par
\bigskip
\noindent {\bf Participants} \hfill 79\par
\bigskip
\noindent {\bf Talk Video Links} \hfill 82

\newpage
\Title{Foreword}
\bigskip\bigskip
\begin{raggedright}  

{\it Michael Way\\
NASA/Goddard Institute for Space Studies\\
2880 Broadway\\
New York, New York, USA}
\bigskip

{\it Catherine Naud\\
Department of Applied Physics and Applied Mathematics\\
Columbia University, New York, New York, USA \\
and\\
NASA/Goddard Institute for Space Studies\\
2880 Broadway\\
New York, New York, USA}

\bigskip\bigskip
\end{raggedright}  

The purpose of the New York Workshop on Computer, Earth and Space
Sciences is to bring together the New York area's finest Astronomers,
Statisticians, Computer Scientists, Space and Earth Scientists to
explore potential synergies between their respective fields. The 2011
edition (CESS2011) was a great success, and we would like to thank
all of the presenters and participants for attending. 

This year was also special as it included authors from the upcoming book titled
``Advances in Machine Learning and Data Mining for Astronomy.''
Over two days, the latest advanced techniques used to analyze the
vast amounts of information now available for the understanding of
our universe and our planet were presented. These proceedings attempt
to provide a small window into what the current state of research
is in this vast interdisciplinary field and we'd like to thank the
speakers who spent the time to contribute to this volume.

This year all of the presentations were video taped and those
presentations have all been uploaded to YouTube for easy access.
As well, the slides from all of the presentations are available
and can be downloaded from the workshop
website\footnote{http://www.giss.nasa/gov/meetings/cess2011}.

We would also like to thank the local NASA/GISS staff for their
assistance in organizing the workshop; in particular Carl Codan
and Patricia Formosa. Thanks also goes to Drs. Jim Hansen and
Larry Travis for supporting the workshop and allowing us to
host it at The Goddard Institute for Space Studies again.

\newpage
\Title{Introduction}
\bigskip\bigskip
\begin{raggedright}  

{\it Michael Way\\
NASA/Goddard Institute for Space Studies\\
2880 Broadway\\
New York, New York, USA}

\bigskip\bigskip
\end{raggedright}  


This is the 2nd time I've co-hosted the New York Workshop on Computer, Earth,
and Space Sciences (CESS). My reason for continuing to do so is that, like
many at this workshop, I'm a strong advocate of interdisciplinary research. My
own research institute (GISS\footnote{Goddard Institute for Space Studies})
has traditionally contained people in the fields of Planetary Science,
Astronomy, Earth Science, Mathematics and Physics. We believe this has
been a recipe for success and hence we also continue partnerships
with the Applied Mathematics and Statistics Departments at Columbia University
and New York Unversity. Our goal with these on-going workshops is to
find new partnerships between people/groups in the entire New York area
who otherwise would never have the opportunity to meet and share ideas
for solving problems of mutual interest.

My own science has greatly benefitted over the years via collaborations
with people I would have never imagined working with 10 years ago.
For example, we have managed to find new ways of using Gaussian Process
Regression (a non-linear regression technique) \citep{Way09} by
working with linear algebra specialists at the San Jose State University
department of Mathematics and Computer Science. This has led to novel
methods for inverting relatively large ($\sim$100,000$\times$100,000) non-sparse
matrices for use with Gaussian Process Regression \citep{foster09}.

As we are all aware, many scientific fields are also dealing with a data
deluge which is often approached by different disciplines in different ways.
A recent issue of
Science Magazine\footnote{http://www.sciencemag.org/site/special/data}
has discussed this in some detail \citep[e.g.][]{Baranuik}. It
has also been discussed in the recent book ``The Fourth Paradigm''
\cite{microsoft}.
What the Science articles made me the most aware of is my own continued
narrow focus. For example, there is a great deal that could be shared between
the people at this workshop and the fields of Biology, Bio-Chemistry, Genomics
and Ecologists to name a few from the Science article. This is
particularly embarrassing for myself since in 2004 I attended
a two-day seminar in Silicon Valley that discussed several chapters in the
book ``The Elements of Statistical Learning'' \citep{hastie}.
Over 90\% of the audience were Bio-Chemists, while I was only
one of two Astronomers.

Another area which I think we can all agree most fields can benefit from is
better (and cheaper) methods for displaying and hence interrogating our data.
Later today I will discuss a program called viewpoints \citep{GLW2010}
which can be used to look at modest sized multivariate data sets
on an individual desktop/laptop. Another of the Science Magazine articles
\citep{fox} discusses a number of ways to look at data in less
expensive way.

In fact several of the speakers at the CESS workshop this year are also
contributors to a book in progress \citep{AMLDM} that has chapters written
by a number of people in the fields of Astronomy, Machine Learning
and Data Mining who have themselves engaged in interdisciplinary
research -- this being one of the rationales for inviting them to
contribute to this volume.

Finally, although I've restricted myself to the ``hard sciences''
we should not forget that interdisciplinary research is taking
place in areas that perhaps only a few of us are familiar with. For example,
I can highly recommend a recent book \citep{morris} that discusses possible
theories for the current
western lead in technological innovation. The author (Ian Morris)
uses data and methodologies from the fields of History, Sociology,
Anthropology/Archaeology, Geology, Geography and Genetics to support the thesis
in the short title of his book: ``Why The West Rules -- For Now''.

Regardless, I would like to thank all of the speakers for coming to New York
and also for contributing to the workshop proceedings.

\newpage

\setcounter{section}{0}
\setcounter{figure}{0}

\newtheorem{prop}{Proposition}[section] 
\newtheorem{thm}{Theorem}[section]              
\Title{On a new approach for estimating threshold crossing times with an
application to global warming}
\bigskip\bigskip
\begin{raggedright}

{\it Victor J. de la Pe\~{n}a\footnote{Joint work with Brown, M., Kushnir, Y., Ravindarath, A. and Sit, T}\\
Columbia University\\
Department of Statistics\\
New York, New York, USA}

\bigskip\bigskip
\end{raggedright}

\section*{Abstract}
Given a range of future projected climate trajectories taken from a multitude of models or scenarios, we attempt to find the best way to determine the threshold crossing time. In particular, we compare the proposed estimators to the more commonly used method of calculating the crossing time from the average of all trajectories (the mean path) and show that the former are superior in different situations. Moreover, using one of the former approaches also allows us to provide a measure of uncertainty as well as other properties of the crossing times distribution. In the cases with infinite first-hitting time, we also provide a new approach for estimating the cross time and show that our methods perform better than the common forecast. As a demonstration of our method, we look at the projected reduction in rainfall in two subtropical regions: the US Southwest and the Mediterranean. \\~\\
KEY WORDS: Climate change; First-hitting time; Threshold-crossing; Probability bounds; Decoupling.

\section*{Introduction: Data and Methods}
The data used to carry out the demonstration of the proposed method are time series of Southwest (U.S.) and Mediterranean region precipitation, calculated from IPCC Fourth Assessment (AR4) model simulations of the twentieth and twenty-first centuries \citep{randall2007}.
To demonstrate the application of our methods, simulated annual mean precipitation time series, area averaged over the US West ($125^{\circ}$W to $95^{\circ}$W and $25^{\circ}$N to $40^{\circ}$N) and the Mediterranean ($30^{\circ}$N to $45^{\circ}$N and $10^{\circ}$W to $50^{\circ}$E), were assembled from nineteen models. Refer to \cite{seager2007}
and the references therein for details. 

\section*{Optimality of an unbiased estimator}
\subsection*{An unbiased estimator}  
Before discussing the two possible estimators, we define $\boldsymbol{X}(t) = \{X_1(t), \ldots, X_n(t)\}$ be the outcomes of $n$ models (stochastic processes in the same probability space). The first hitting time of the $i$th simulated path $X_i$ with $T$ bounded is defined as 
$$T_{r,i} := \inf\left\{t\in[0, \tau]: X_{i}(t) \geq r\right\} \text{ where } X_i(t) \geq 0~~~~,~~~i=1, \ldots, n(=19).$$
Unless otherwise known, we assume that the paths are equally likely to be close to the ``true'' path. Therefore, we let 
\begin{equation}
T_r = T_j \text{ with probability } \frac{1}{n}~~~~,~~~~j=1, \ldots, n(=19),
\label{eq:assumption}
\end{equation}
where $T_r$ denotes the true path. There are two possible ways to estimate the first-hitting time of the true path, namely 
\begin{enumerate}
         \item Mean of the first-hitting time: 
         \begin{equation*} 
         T^{(UF)}_r := \frac{1}{n}\sum^{n}_{i=1}T_{r,i}, 
         \label{est:nf}
         \end{equation*}
         \item First-hitting time of the mean path: 
         \begin{equation*} 
         T^{(CF)}_r := \inf\left\{t\in[0, \tau]: \bar{X}_n(t) := \frac{1}{n}\sum^{n}_{i=1}X_{i}(t) \geq r\right\}.
         \label{est:cf}
         \end{equation*}
         \end{enumerate}
\begin{prop}
The unbiased estimator $T^{(UF)}_b$ outperforms the traditional estimator $T^{(CF)}_b$ 
in terms of (i) mean-squared error and (ii) Brier skill score.  $a^{-1}_{(n)}(r)$, to be specified in Theorem 3.1, is preferred in cases where $T^{(UF)}_r = \infty$.\\~\\
Remark: By considering the crossing times of individual paths, we can obtain an empirical CDF for $T_r$, which is useful for modeling various statistical properties of $T_r$.           
\end{prop}

\subsection*{Extending boundary crossing of non-random functions to that of stochastic processes}
In the situations in which not all the simulated paths cross the boundary before the end of the experiment, we propose a remedy which can be summarized in the following theorem. For details, refer to \cite{brown2011} 
\begin{thm}
Let $X_s \geq 0$, $a_{(n)}(t) = E\sup_{s\leq t}X_s = n^{-1}\sum^n_{i=1}\sup_{s\leq t}Y_{s,i}$. Assume $a_n(t)$ is increasing (we can also use a generalized inverse) with $a_{(n)}^{-1}(r) = t_r = \inf\{t > 0: a_{(n)}(t) = r\} \longrightarrow a(t)$, we can obtain bounds, under certain conditions:
        \begin{equation*}
        \frac{1}{2}a_{(n)}^{-1}(r/2) \leq E[T_r] \leq 2 a_{(n)}^{-1}(r),
        \end{equation*} 
Remark: The lower bound is universal.       
\end{thm}
    \begin{figure}[h]
    \begin{center}
      \includegraphics[width=12cm]{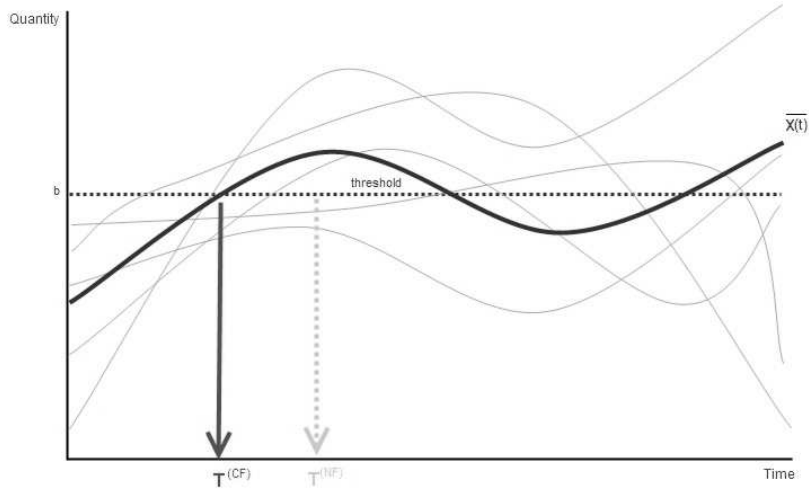}
      \caption{Illustrating how to obtain $T^{(UF)}$ and $T^{(CF)}$}
    \end{center}
    \end{figure}

    \begin{figure}[h]
    \begin{center}
      \includegraphics[width=12cm]{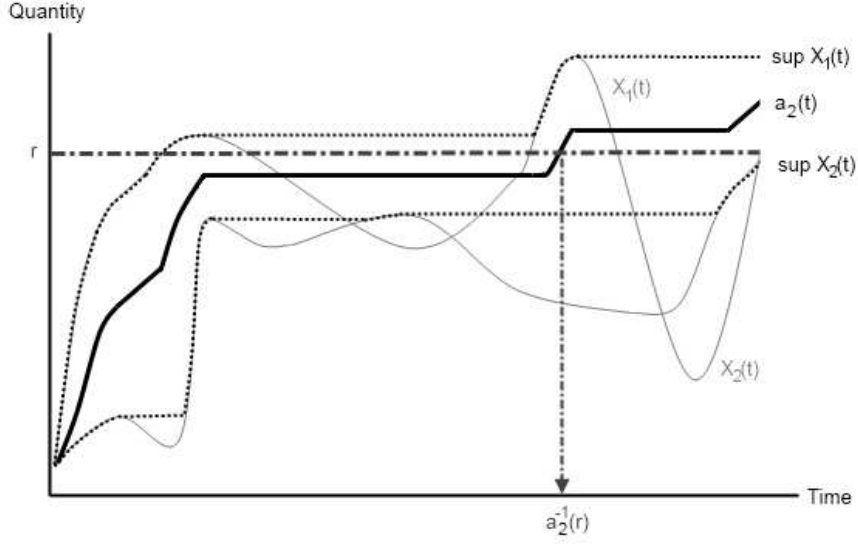}
      \caption{Illustration of $a_{(n)}(t)$ and $a_{(n)}^{-1}(r)$.}
    \end{center}
    \end{figure}

\newpage
\section*{Results}
Details of the results are tabulated as follows:
\begin{center}
       \begin{tabular}{| c | c | c | c | c |}
       \hline
        & $T^{(UF)} := \sum n^{-1}T_i$& $ T^{(CF)}(n^{-1})$ & $\hat{M}$ & $a^{-1}_{(n)}(r)$\\\hline
       Mediterranean & 2010.21 & 2035 & 2018 & 2008\\\hline
       Southwest US &  $\infty$ & 2018 & 2011 &  2004\\\hline
       \end{tabular}
       \end{center}
       \footnotesize where $T^{(UF)}$, $T^{(CF)}$, $\hat{M}$ and $a^{-1}_{(n)}(r)$ denote respectively the mean hitting times of the simulated paths, the hitting time of the mean simulated path, the median hitting times of the simulated paths and the hitting time estimate based on Theorem 3.1. 19 paths are simulated for both Mediterranean and Southwest US regions. The infinity value for the Southwest US region is due to the fact that there are three paths that do not cross the boundary. If we just include the paths that cross the boundary, we will have $T^{(UF)} = 2004.63$. Clearly, in the case of Southwest,$a_(n)^{-1}(r)$ is better than $T_r$ wich has infinite expectation.\normalsize~\\
       
       According to the current estimates, the drought in the Southwest region is already in process. This observation shows a case where $T^{(UF)}$ and $a^{-1}_{(n)}(r)$ provide better forecasts than $T^{(CF)}$ or the median.

\newpage
\setcounter{section}{0}
\setcounter{figure}{0}

\Title{Cosmology through the large-scale structure of the Universe}
\bigskip\bigskip
\begin{raggedright}  

{\it Eyal A. Kazin\footnote{eyalkazin@gmail.com}\\
Center for Cosmology and Particle Physics\\
New York University, 4 Washington Place\\
New York, NY 10003, USA}
\end{raggedright}  

\section*{Abstract}

The distribution of matter contains a lot of cosmological information. 
Applying N-point statistics one can measure the geometry and expansion 
of the cosmos as well as test General Relativity at scales of millions
to billions of light years.  In particular, I will discuss an exciting
recent measurement dubbed the ``baryonic acoustic feature", which has
recently been detected in the Sloan Digital Sky Survey galaxy sample.
It is the largest known ``standard ruler" (half a billion light years
across), and is being used to investigate the nature of the acceleration
of the Universe.

\section*{The questions posed by $\Lambda$CDM}

The Cosmic Microwave Background (CMB) shows us 
a picture of the early Universe which was very uniform \citep{penzias65a}, 
yet with enough inhomogeneities \citep{smoot92a} to seed the 
structure we see today in the form of galaxies and 
the cosmic-web. 
Ongoing sky surveys are 
measuring deeper into the Universe with 
high edge technology
transforming cosmology into 
a precision science.

The leading ``Big Bang" model today 
is dubbed $\Lambda$CDM.
While shown to be superbly consistent with many 
independent astronomical probes, 
it indicates that the ``regular" material 
(atoms, radiation) comprise of 
only $5\%$ of the energy budget, 
hence challenging our current understanding of physics. 

The $\Lambda$ is a reintroduction of 
Einstein's so-called cosmological constant. 
He originally introduced it to 
stabilize a Universe that could expand or contract, 
according to General Relativity.
At present, it is considered a mysterious energy with a repulsive 
force that explains the acceleration of the observed Universe. 
This acceleration was first noticed 
through super-novae distance-redshift relationships 
\citep{riess98, perlmutter99a}. 
Often called {\it dark energy}, 
it has no clear explanation, 
and most cosmologists would happily 
do away with it, once 
a better intuitive explanation emerges. 
One stream of thought is modifying 
General Relativity on very large scales, e.g, 
by generalizing to higher dimensions. 

Cold dark matter (CDM), 
on the other hand, 
has gained its ``street-cred" 
throughout recent decades, 
as an invisible substance 
(meaning not interacting with radiation), 
but seen time and time again 
as the dominant gravitational 
source.  
Dark matter is required 
to explain various measurements 
as the 
virial motions of galaxies within clusters
\citep{zwicky33a}, 
the rotation curves of galaxies \citep{rubin70a}, 
the gravitational lensing of background galaxies, 
and collisions of galaxy clusters \citep{clowe04a}.
We have yet to detect dark matter 
on Earth, although 
there already have 
been false positives. 
Physicists  hope to 
see convincing evidence 
emerge from the 
Large Hadron Collider 
which is bashing protons 
at near the speed of light. 

One of the most convincing pieces 
of evidence for dark matter is 
the growth of the large-scale structure of the 
Universe, the subject of this essay. 
The CMB gives us a picture of the Universe
when it was 
 one thousand time 
smaller than present. 
Early Universe inhomogeneities seen through 
temperature fluctuations in the CMB 
are of the order one part in $10^5$. 
By measuring the distribution 
of galaxies, the 
structure in the recent Universe is probed  to 
percent level at scales of 
hundreds of millions of light-year scales 
and  it can also be probed at the 
unity level  and higher at ``smaller" cosmic scales 
of thousands of light-years. 
These tantalizing differences in structure 
can not be explained by  the gravitational attraction of regular material 
alone (atoms, molecules, stars, galaxies etc.),  
but  can be explained with non-relativistic dark matter. 
Similar arguments show 
that the dark matter consists 
of $\sim 20\%$ of the energy 
budget, and dark energy $\sim 75\%$. 

The distribution of matter, hence, 
is a vital test for any cosmological model. 

\section*{Acoustic oscillations as a cosmic ruler}
Recently an important feature dubbed 
the {\it  baryonic acoustic feature} has been 
detected in galaxy clustering 
\citep{eisenstein05b, percival09b, kazin10a}. 
The feature has 
been detected significantly 
in the anisotropies of the CMB 
by various Earth and space 
based missions 
\citep[e.g.][]{torbet99a, komatsu09a}. 
Hence, cosmologists have 
made an important connection 
between the early and late Universe. 

When the Universe was 
much smaller than today,  
energetic radiation dominated 
and did not enable the formation of atoms. 
Photon pressure  
on the free electrons and 
protons (collectively called {\it baryons}), 
 caused  them to  propagate 
as a fluid in acoustic wave fashion. 
A useful analogy to have in mind 
is a pebble dropped in water perturbing 
 it and forming a wave. 

As the Universe expanded it cooled down and the first 
atoms formed freeing the radiation, 
which we now measure as the CMB. 
Imagine the pond freezing,  
including the wave. 
As the atoms are no longer being pushed 
they slow down, 
and are now gravitationally 
bound to dark matter. 

This means that around 
every over density, where 
the plasma-photon waves (or pebble) 
originated, we  
expect  an excess of material 
at a characteristic 
radius 
of the wave when it froze, 
dubbed the {\it sound horizon}. 

\begin{figure}[htp]
\includegraphics[width=\textwidth] {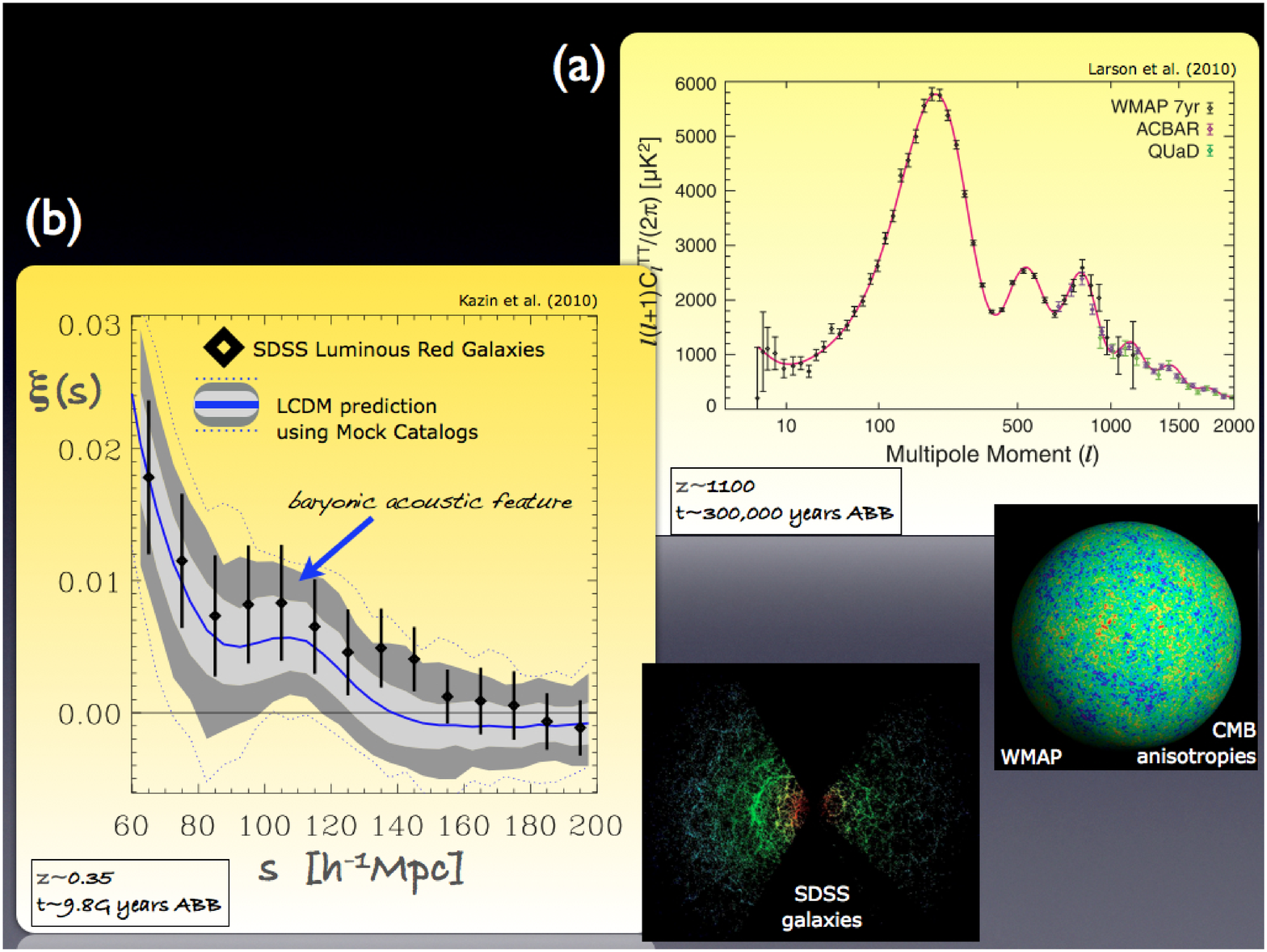} 
\vspace{0.5in}
\caption{
The baryonic acoustic feature in the large-scale structure of the Universe. 
The solid lines are $\Lambda$CDM predictions. 
(a) Temperature fluctuations in the CMB (2D projected $k$-space), 
measured by WMAP, ACBAR and QUaD. 
The feature is the form of peaks and troughs, 
and is detected to very high significance. 
These anisotropies  are at the level of $10^{-5}$, 
and reflect the Universe as it was 1000-fold smaller 
than today ($\sim 13.3$ billion year ago). 
(b) SDSS luminous galaxy 3D clustering $\xi$  at very large scales 
($100$\hmpc corresponds to $\sim 0.5$ billion light years).  
The feature is detected consistent with predictions. 
Notice $\xi$  is of order $1\%$ at the feature, 
showing a picture of the Universe $\sim 3$ billion years ago. 
The gray regions indicate $68, \ 95\%$ CL regions of simulated  
mock galaxy catalogs, reflecting cosmic variance. These 
will be substantially reduced in the future with larger volume 
surveys. (ABB means time ``after big bang")
}
\label{baf_plot}
\end{figure}

In practice, this does not 
happen in a unique place, 
but throughout the whole 
Universe 
(think of throwing many pebbles into the pond). 
This means that 
we expect 
to measure a characteristic 
correlation length 
in the anisotropies of the CMB, 
as well as in the clustering of matter 
in a statistical manner. 
Figure \ref{baf_plot}  demonstrates 
the detection of the feature 
in the CMB temperature anisotropies 
\citep{larson11a}
and in the clustering 
of luminous red galaxies 
\citep{eisenstein01a}. 

As mentioned before, 
the $\sim 10^5$ increase 
in the amplitude of the 
inhomogeneities 
between early (CMB) and late 
Universe (galaxies) is explained 
very well with dark matter. 
The height of the 
baryonic acoustic feature 
also serves as a firm 
prediction of the CDM paradigm. 
If there was no dark matter, 
the relative amplitude 
of the feature would be much 
higher. 
An interesting anecdote 
is that we happen to live 
in an era when the feature 
is still detectable in 
galaxy clustering. 
Billions of years from now, 
it will be washed away,  
due to gravitational interplay between 
dark matter and galaxies. 

In a practical sense, as the feature 
spans a 
characteristic scale,  
it can be used as a cosmic ruler. 
The signature in the 
anisotropies of the CMB 
(Figure \ref{baf_plot}a), 
calibrates this ruler by 
measuring the sound-horizon 
currently to an accuracy of $\sim 1.5\%$ \citep{komatsu09a}.

By measuring the feature in galaxy clustering transverse 
to the line-of-sight, you can think of it 
as the base of a triangle, for which we 
know the observed angle, and hence 
can infer the distance to the galaxy sample. 
Clustering along the line-of-sight 
is an even more powerful measurement, 
as it is sensitive to the expansion of the Universe.
By measuring expansion rates 
one can test effects of dark energy.
Current measurements 
show that the baryonic acoustic feature 
in Figure \ref{baf_plot}b,
 can 
be used to measure 
the distance to $\sim 3.5$ 
billion light-years  
to an accuracy of  $\sim 4\%$ \citep{percival09b, kazin10a}.

\section*{Clustering- the technical details} 
As dark matter can not be seen directly, 
luminous objects, as galaxies, 
can serve as tracers, like the tips of icebergs. 
Galaxies are thought to form 
in regions of high dark matter density.
An effective way 
to measure galaxy clustering 
(and hence inferring the matter distribution)  
is through two-point correlations of 
over-densities.  

An over-density at point $\vec{x}$ is 
defined as the contrast to the mean density 
$\overline{\rho}$:
\beq
\delta(\vec{x})\equiv\frac{\rho(\vec{x})}{\overline{\rho}}-1. 
\eeq
The auto-correlation function, defined as the 
joint probability of measuring an excess of 
density at a given separation $r$ is defined as:
\beq
\xi(r)\equiv \avg{\delta(\vec{x})\delta(\vec{x}+\vec{r})}, 
\eeq 
where the average is over the volume, and 
the cosmological principle assumes 
statistical isotropy. 
This is related to the Fourier complementary 
power spectrum P($k$). 

For P$(k)$, it is common to smooth 
out the galaxies into density fields, 
Fourier transforming $\delta$ and 
convolving with a ``window function" 
that describes the actual geometry of the survey.

The estimated $\xi$, in practice, 
is calculated by counting galaxy pairs:
\beq
\hat{\xi}(r)=\frac{ DD(r) }{ RR(r) }-1, 
\eeq
where $DD(r)$ is the normalized number 
of galaxy pairs within a spherical 
shells of radius 
$r \pm \frac{1}{2}\Delta r$. 
This is compared to random points distributed 
according to the survey geometry, 
where $RR$ is the random-random 
normalized pair count.  
By normalized I refer to the fact that 
one uses many more random points than 
data points to reduce Poisson shot noise. 
\cite{landy93a} show that an estimator 
that minimizes the variance is: 
\beq
\hat{\xi}(r)=\frac{ DD(r) + RR(r) - 2DR(r)}{ RR(r) }, 
\eeq
where $DR$ are the normalized data-random pairs.

\section*{The Sloan Digital Sky Survey}
Using a dedicated $2.5$ meter 
telescope, the SDSS has industrialized 
(in a positive way!) astronomy. 
In January 2011,  
they publicly released 
an image of 
one third of  the sky,   
and detected  
$469$ million objects from astroids 
to galaxies\footnote{http://www.sdss3.org/dr8/}
\citep{aihara11a}. 

These images give a 2D projected 
image of the Universe. 
This is followed up by targeting 
objects of interest, 
obtaining their spectroscopy. 
The spectra contains information 
about the composition of the objects.
As galaxies and quasars 
have signature spectra, 
these can be used as a templates 
to measure the Doppler-shift. 
The expanding Universe causes 
these to be redshifted. 
The redshift $z$ can be simply 
related to the distance $d$ through 
the Hubble equation  at low $z$:
\beq\label{hubble_eq}
cz=Hd,
\eeq
where $c$ is the speed of light 
and the Hubble parameter 
 $H$ [1/time]
is the expansion rate of the 
Universe. 
Hence, by measuring $z$, 
observers obtain a 3D 
picture of the Universe, 
which can be used to measure clustering. 
Dark energy effects 
Equation \ref{hubble_eq} 
through $H(z)$, when generalizing 
for larger distances. 

The SDSS team has obtained 
spectroscopic redshifts of over 
a million 
objects in the largest volume 
to date. 
It is now in its third phase, 
obtaining more spectra for  
various missions including:
 improving measurements  
of the baryonic acoustic feature 
(and hence measuring dark energy) 
by measuring a larger and deeper volume, 
learning the structure of the Milky Way, 
and detection of exoplanets 
\citep{eisenstein11a}.

\section*{Summary}
Cosmologists are showing 
that there is much more than 
meets the eye. 
It is just a matter of 
time until dark matter will 
be understood, 
and might I be bold enough 
to say harnessed? 
The acceleration of the Universe, 
is still a profound mystery, 
but equipped with tools such as the 
baryonic acoustic feature, 
cosmologists will 
be able to provide 
rigorous tests. 

\bigskip\bigskip

E.K was partially supported by a Google Research Award and NASA Award
NNX09AC85G.

\newpage
\setcounter{section}{0}
\setcounter{figure}{0}

\Title{On the Shoulders of Gauss, Bessel, and Poisson: Links, Chunks, Spheres,
and Conditional Models}
\bigskip\bigskip
\begin{raggedright}  
{\it William D Heavlin \\
Google, Inc.\\
Mountain View, California, USA}
\bigskip\bigskip
\end{raggedright}

\section*{Abstract}
We consider generalized linear models (GLMs) and the associated
exponential family (``links''). Our data structure partitions the data
into mutually exclusively subsets (``chunks''). The conditional likelihood
is defined as conditional on the within-chunk histogram of the
response. These likelihoods have combinatorial complexity. To compute
such likelihoods efficiently, we replace a sum over permutations
with an integration over the orthogonal or rotation group (``spheres'').
The resulting approximate likelihood gives rise to estimates that are
highly linearized, therefore computationally attractive. Further, this
approach refines our understanding of GLMs in several directions.

\section*{Notation and Model}

Our observations are chunked into subsets indexed
by $g:$ \emph{$(y_{gi},\mathbf{x}_{gi}:\; g=1,2,\ldots,G;\: i=1,\ldots,n_{g})$.}
The \emph{g-}th chunk's responses are denoted by
$\mathbf{y}_{g}=(y_{g1},y_{g2},\ldots,y_{gn_{g}})$
and its feature matrix by $\mathbf{X}_{g};$ its \emph{i}-th row is
$\mathbf{x}_{gi}^{\mathbf{T}}.$ Our framework is that of the generalized
linear model \citep{MN1999}:

\begin{equation}
\Pr\{y_{gi}|\mathbf{x}_{gi}^{\mathbf{T}}\mathbf{\beta}\}=\exp\{y_{gi}\mathbf{x}_{gi}^{\mathbf{T}}\mathbf{\beta})+h_{1}(y_{gi})+h_{2}(\mathbf{x}_{gi}^{\mathbf{T}}\mathbf{\beta})\}.\label{eq:1}
\end{equation}

\section*{The Spherical Approximation}

Motivated by the risk of attenuation, we condition ultimately on the
variance of $\mathbf{y}_{g}.$ The resulting likelihood consists of
these terms, indexed by $g:$

\begin{equation}
\exp\{L_{cg}(\mathbf{\beta})\}\approx\frac{\exp\{\mathbf{y}_{g}^{\mathbf{T}}\mathbf{X}_{g}\mathbf{\beta}\}}{\mbox{ave}\ensuremath{\{\exp\{\mathbf{y}_{g}^{\mathbf{T}}\mathbf{P}_{\tau}^{\mathbf{T}}\mathbf{X}_{g}\mathbf{\beta}\}|\tau\in\mbox{orthogonal}\}}}\label{eq:2}
\end{equation}

Free of intercept terms, this likelihood resists attenuation. The
rightmost term of (\ref{eq:2}) reduces to the von Mises-Fisher
distribution \citep{MJ2000,WW1956} and is computationally attractive \citep{P2010}. 

\begin{figure}
\includegraphics[scale=1.0]{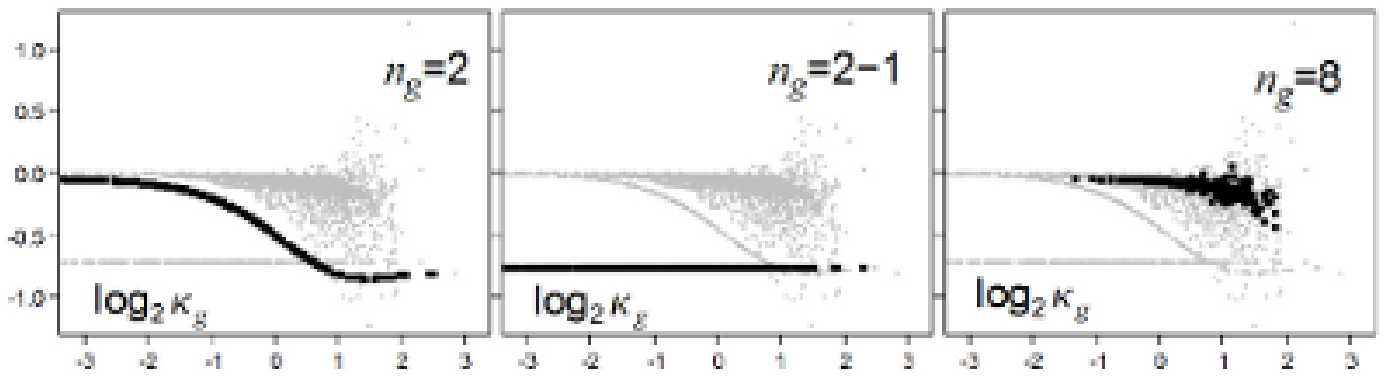}
\caption{\emph{Numerically calculated values of $\frac{\partial}{\partial\kappa}\log Q$
as a function of radius $\kappa.$}}
\label{heavlin-fig01}
\end{figure}

Figure \ref{heavlin-fig01} assesses the spherical approximation. The x-axis
is the radius $\kappa=||\mathbf{y}_{g}||\times||\mathbf{X}_{g}\mathbf{\beta}||$,
the y-axis the differential effect of equation (\ref{eq:2})'s two
denominators. Panel (c) illustrates how larger chunk sizes $n_{g}$
improve the spherical approximation. Panel (a) and (b) illustrates
how the approximation for $n_{g}=2$ can be improved by a continuity
correction.

\section*{Some Normal Equations}

From (\ref{eq:2}) these maximum likelihood equations follow:

\begin{equation}
[\sum_{g}\frac{\rho_{g}}{r_{g}}\mathbf{X}_{g}^{\mathbf{T}}\mathbf{X}_{g}]\mathbf{\hat{\beta}}=\mathbf{X}_{g}^{\mathbf{T}}\mathbf{y}_{g},\label{eq:3}
\end{equation}
which are nearly the same as those of Gauss. Added is the ratio $\rho_{g}/r_{g},$
which throttles chunks with less information; to first order, it equals
the within-chunk variance. 

The dependence of $\rho_{g}/r_{g}$ on $\mathbf{\beta}$ is weak,
so the convergence of (\ref{eq:3}) is rapid. Equation (\ref{eq:3})
resembles iteratively reweighted least squares \citep{J2006}, but is more
attractive computationally. To estimate many more features, we investigate
marginal regression \citep{FL2008} and boosting \citep{SS1999}.

Conditional models like those in (\ref{eq:2}) do not furnish estimates
of intercepts. The theory of conditional models therefore establishes
a framework for multiple-stage modeling.

\newpage

\setcounter{section}{0}
\setcounter{figure}{0}

\Title{Mining Citizen Science Data: Machine Learning Challenges}
\bigskip\bigskip
\begin{raggedright}  

{\it Kirk Borne\\
School of Physics, Astronomy \& Computational Science\\
George Mason University
Fairfax, Virginia, USA}

\bigskip\bigskip
\end{raggedright}  

Large sky surveys in astronomy, with their open data policies (``data for 
all'') and their uniformly calibrated scientific databases, are key 
cyberinfrastructure for astronomical research.  These sky survey 
databases are also a major content provider for educators and the general 
public.  Depending on the audience, we recognize three broad modes of 
interaction with sky survey data (including the image archives and the 
science database catalogs).  These modes of interaction span the 
progression from information-gathering to active engagement to discovery.  
They are:
\begin{itemize}
\item[a.)] Data Discovery -- What was observed, when, and by whom?  Retrieve 
observation parameters from an sky survey catalog database. Retrieve 
parameters for interesting objects.
\item[b.)] Data Browse -- Retrieve images from a sky survey image archive. View 
thumbnails. Select data format (JPEG, Google Sky KML, FITS).  Pan 
the sky and examine catalog-provided tags (Google Sky, World Wide 
Telescope).
\item[c.)] Data Immersion -- Perform data analysis, mining, and visualization. 
Report discoveries. Comment on observations. Contribute followup 
observations.  Engage in social networking, annotation, and tagging.  
Provide classifications of complex images, data correlations, data 
clusters, or novel (outlying, anomalous) detections.
\end{itemize}

In the latter category are Citizen Science research experiences.  The world 
of Citizen Science is blossoming in many ways, including century-old 
programs such as the Audubon Society bird counts and the American 
Association of Variable Star Observers (at aavso.org) continuous 
monitoring, measurement, collation, and dissemination of brightness 
variations of thousands of variable stars, but now including numerous 
projects in modern astronomy, climate science, biodiversity, watershed 
monitoring, space science, and more. The most famous and successful of 
these is the Galaxy Zoo project (at galaxyzoo.org), which is ``staffed'' by 
approximately 400,000 volunteer contributors.  Modern Citizen Science 
experiences are naturally online, taking advantage of Web 2.0 
technologies, for database-image-tagging mash-ups.  It takes the form of 
crowd-sourcing the various stages of the scientific process.  Citizen 
Scientists assist scientists' research efforts by collecting, organizing, 
characterizing, annotating, and/or analyzing data. Citizen Science is one 
approach to engaging the public in authentic scientific research 
experiences with large astronomical sky survey databases and image 
archives. 

Citizen Science is a term used for scientific research projects in which 
individual (non-scientist) volunteers (with little or no scientific training) 
perform or manage research-related tasks such as observation, 
measurement, or computation.  In the Galaxy Zoo project, volunteers are 
asked to click on various pre-defined tags that describe the observable 
features in galaxy images -- nearly one million such images from the SDSS 
(Sloan Digital Sky Survey, at sdss.org).  Every one of these million 
galaxies has now been classified by Zoo volunteers approximately 200 
times each.  These tag data are a rich source of information about the 
galaxies, about human-computer interactions, about cognitive science, and 
about the Universe.  The galaxy classifications are being used by 
astronomers to understand the dynamics, structure, and evolution of 
galaxies through cosmic time, and thereby used to understand the origin, 
state, and ultimate fate of our Universe.  This illustrates some of the 
primary characteristics (and required features) of Citizen Science: that the 
experience must be engaging, must work with real scientific data, must not 
be busy-work, must address authentic science research questions that are 
beyond the capacity of science teams and computational processing 
pipelines, and must involve the scientists. The latter two points are 
demonstrated (and proven) by: (a) the sheer enormous number of galaxies 
to be classified is beyond the scope of the scientist teams, plus the 
complexity of the classification problem is beyond the capabilities of 
computational algorithms, primarily because the classification process is 
strongly based upon human recognition of complex patterns in the images, 
thereby requiring ``eyes on the data''; and (b) approximately 20 peer-
reviewed journal articles have already been produced from the Galaxy Zoo 
results -- many of these papers contain Zoo volunteers as co-authors, and 
at least one of the papers includes no professional scientists as authors.
The next major step in astronomical Citizen Science (but also including 
other scientific disciplines) is the Zooniverse project (at zooniverse.org).  
The Zooniverse is a framework for new Citizen Science projects, thereby 
enabling any science team to make use of the framework for their own 
projects with minimal effort and development activity.  Currently active 
Zooniverse projects include Galaxy Zoo II, Galaxy Merger Zoo, the Milky 
Way Project, Supernova Search, Planet Hunters, Solar Storm Watch, 
Moon Zoo, and Old Weather.  All of these depend on the power of human 
cognition (i.e., human computation), which is superb at finding patterns in 
data, at describing (characterizing) the data, and at finding anomalies (i.e., 
unusual features) in data. The most exciting example of this was the 
discovery of Hanny's Voorwerp (Figure 1). 
A key component of the Zooniverse research program is the mining of the 
volunteer tags. These tag databases themselves represent a major source 
of data for knowledge discovery, pattern detection, and trend analysis.  We 
are developing and applying machine learning algorithms to the scientific 
discovery process with these tag databases.  Specifically, we are 
addressing the question: how do the volunteer-contributed tags, labels, 
and annotations correlate with the scientist-measured science parameters 
(generated by automated pipelines and stored in project databases)? The 
ultimate goal will be to train the automated data pipelines in future sky 
surveys with improved classification algorithms, for better identification of 
anomalies, and with fewer classification errors.  These improvements will 
be based upon millions of training examples provided by the Citizen 
Scientists.  These improvements will be absolutely essential for projects 
like the future LSST (Large Synoptic Survey Telescope, at lsst.org), since 
LSST will measure properties for at least 100 times more galaxies and 100 
times more stars than SDSS.  Also, LSST will do repeated imaging of the 
sky over its 10-year project duration, so that each of the roughly 50 billion 
objects observed by LSST will have approximately 1000 separate 
observations.  These 50 trillion time series data points will provide an 
enormous opportunity for Citizen Scientists to explore time series (i.e., 
object light curves) to discover all types of rare phenomena, rare objects, 
rare classes, and new objects, classes, and sub-classes. The contributions 
of human participants may include: characterization of countless light 
curves; human-assisted search for best-fit models of rotating asteroids 
(including shapes, spin periods, and varying surface reflection properties); 
discovery of sub-patterns of variability in known variable stars; discovery of 
interesting objects in the environments around variable objects; discovery 
of associations among multiple variable and/or moving objects in a field; 
and more.  

\begin{figure}[!htb]
\includegraphics[scale=0.85]{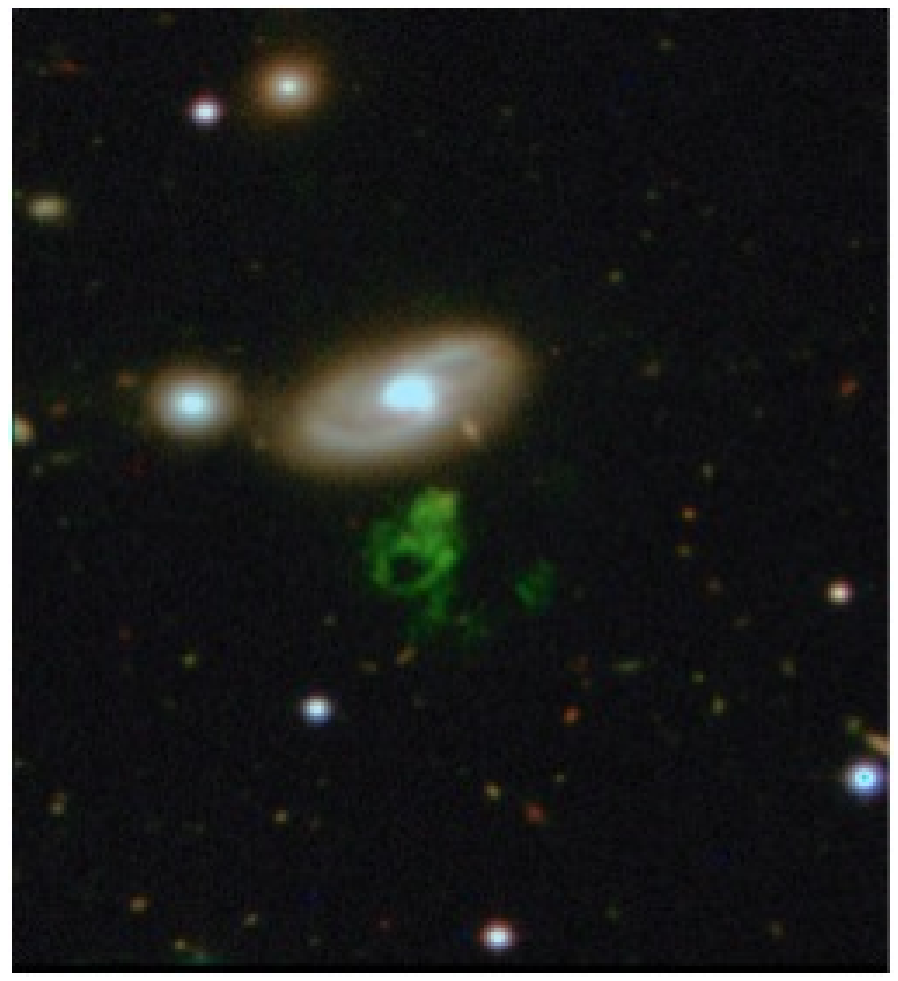}
\includegraphics[scale=0.85]{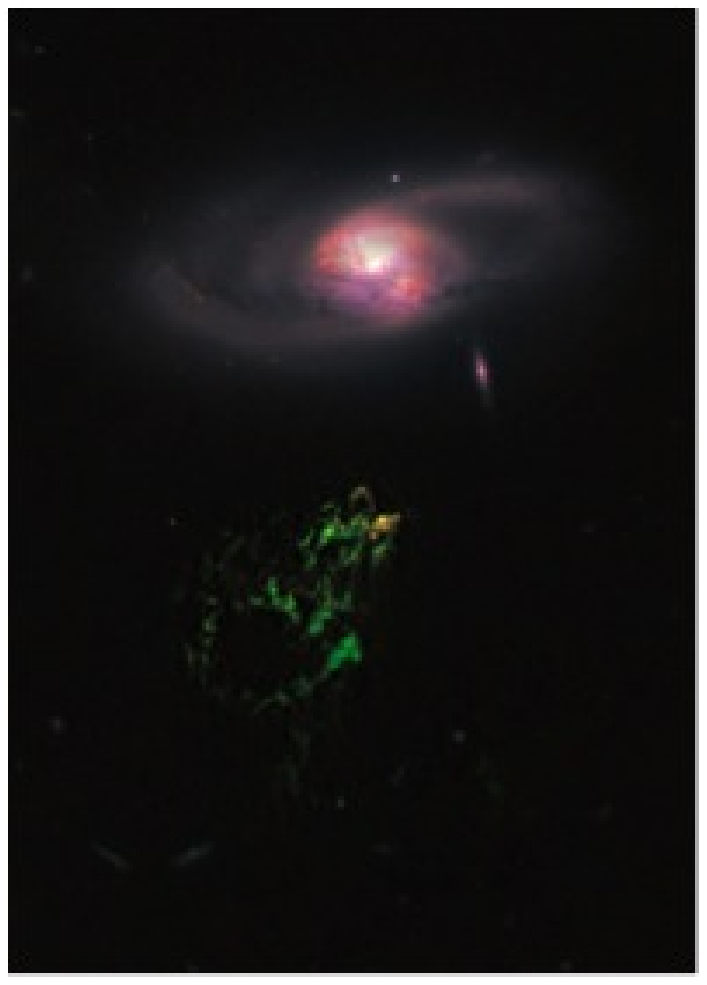}
\caption{Hanny's Voorwerp (Hanny's Object) -- The green gas cloud seen below 
the spiral galaxy in these images was first recognized as something unusual and 
``out of the ordinary'' by Galaxy Zoo volunteer Hanny van Arkel, a Dutch school 
teacher, who was initially focused on classifying the dominant spiral galaxy above 
the green blob.  This object is an illuminated gas cloud, glowing in the emission of 
ionized oxygen.  It is probably the light echo from a dead quasar that was 
luminous at the center of the spiral galaxy about 100,000 years ago.  These 
images are approximately true color. The left image was taken with a ground-
based telescope, and the right image was obtained by the Hubble Space 
Telescope (courtesy W. Keel, the Galaxy Zoo team, NASA, and ESA).}
\label{borne-fig1}
\end{figure}

As an example of machine learning the tag data, a preliminary study by
\citep{baehr}  of 
the galaxy mergers found in the Galaxy Zoo I project was carried out. We 
found specific parameters in the SDSS science database that correlate 
best with ``mergerness'' versus ``non-mergerness''.  These database 
parameters are therefore useful in distinguishing normal (undisturbed) 
galaxies from abnormal (merging, colliding, interacting, disturbed) 
galaxies.  Such results may consequently be applied to future sky surveys 
(e.g., LSST), to improve the automatic (machine-based) classification 
algorithms for colliding and merging galaxies.  All of this was made 
possible by the fact that the galaxy classifications provided by Galaxy Zoo 
I participants led to the creation of the largest pure set of colliding and 
merging galaxies yet to be compiled for use by astronomers.

\newpage
\setcounter{section}{0}
\setcounter{figure}{0}

\Title{Tracking Climate Models\footnote{This is an excerpt from a journal
paper currently under review. The conference version appeared at the NASA
Conference on Intelligent Data Understanding, 2010 \citep{mss10c}}}
\bigskip\bigskip

\begin{raggedright}  
{\it Claire Monteleoni \footnote{cmontel@ccls.columbia.edu}\\
Center for Computational Learning Systems, Columbia University\\
New York, New York, USA}
\bigskip

{\it Gavin A. Schmidt\\
NASA Goddard Institute for Space Studies, 2880 Broadway\\
and\\
Center for Climate Systems Research, Columbia University\\
New York, New York, USA}
\bigskip

{\it Shailesh Saroha\\
Department of Computer Science\\
Columbia University\\
New York, New York, USA}
\bigskip

{\it Eva Asplund\\
Department of Computer Science, Columbia University\\
and\\
Barnard College\\
New York, New York, USA}

\bigskip\bigskip
\end{raggedright}

Climate models are complex mathematical models designed by meteorologists,
geophysicists, and climate scientists, and run as computer simulations, to
predict climate. There is currently high variance among the predictions of
20 global climate models, from various laboratories around the world,
that inform the Intergovernmental Panel on Climate Change (IPCC).  Given
temperature predictions from 20  IPCC global climate models, and over 100 years
of historical temperature data, we track the changing sequence of which model
currently predicts best.  We use an algorithm due to \cite{MJ03},
that models the sequence of observations using a hierarchical learner,
based on a set of generalized Hidden Markov Models, where the identity of the
current best climate model is the hidden variable.  The transition probabilities
between climate models are learned online, simultaneous to tracking the
temperature predictions.  

\begin{figure}
\begin{tabular}{cc}
\includegraphics[trim=1in 0 0 0, clip, width=5.5in]{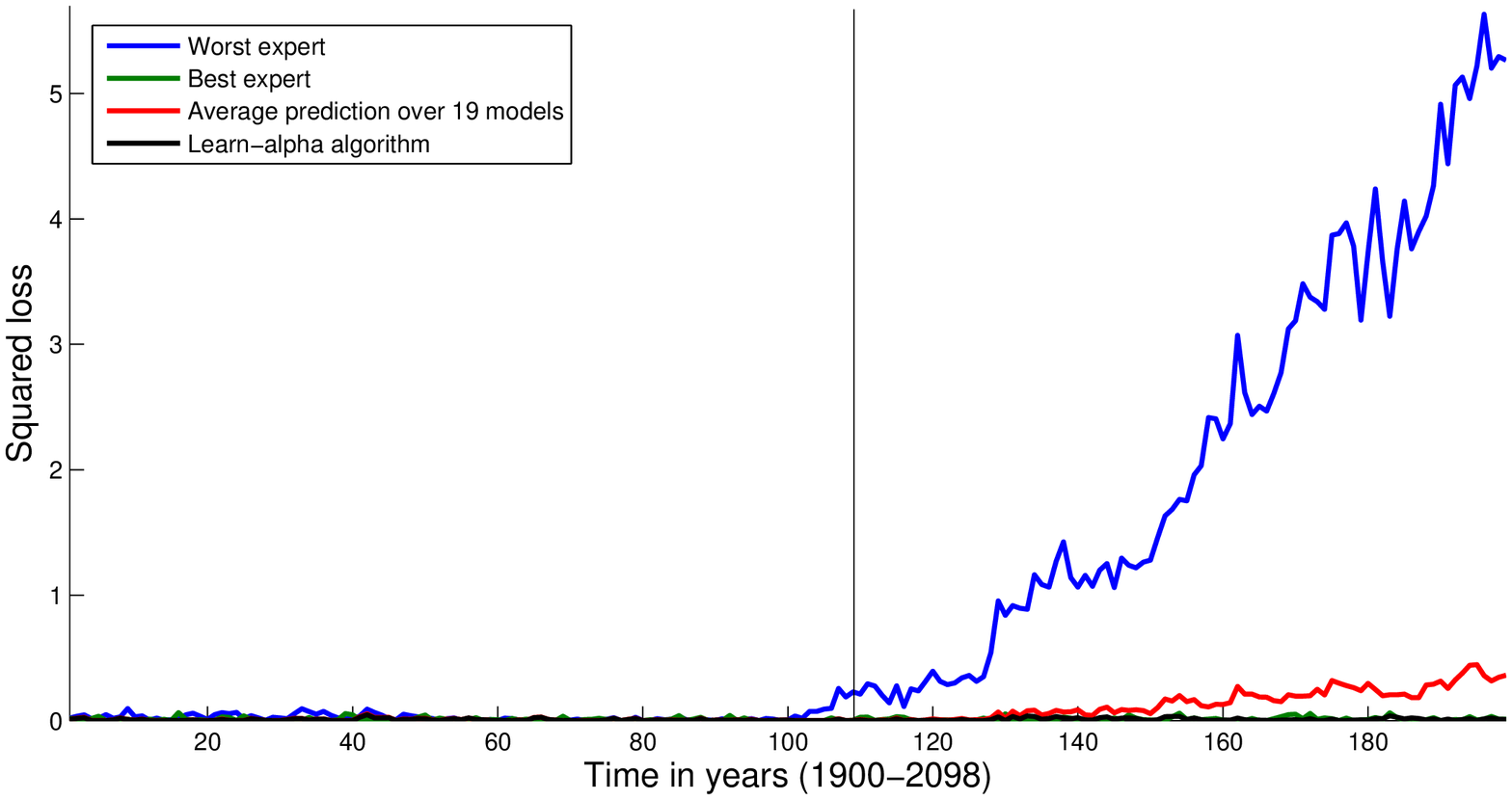} \\
\includegraphics[trim=1in 0 0 0, clip, width=5.5in]{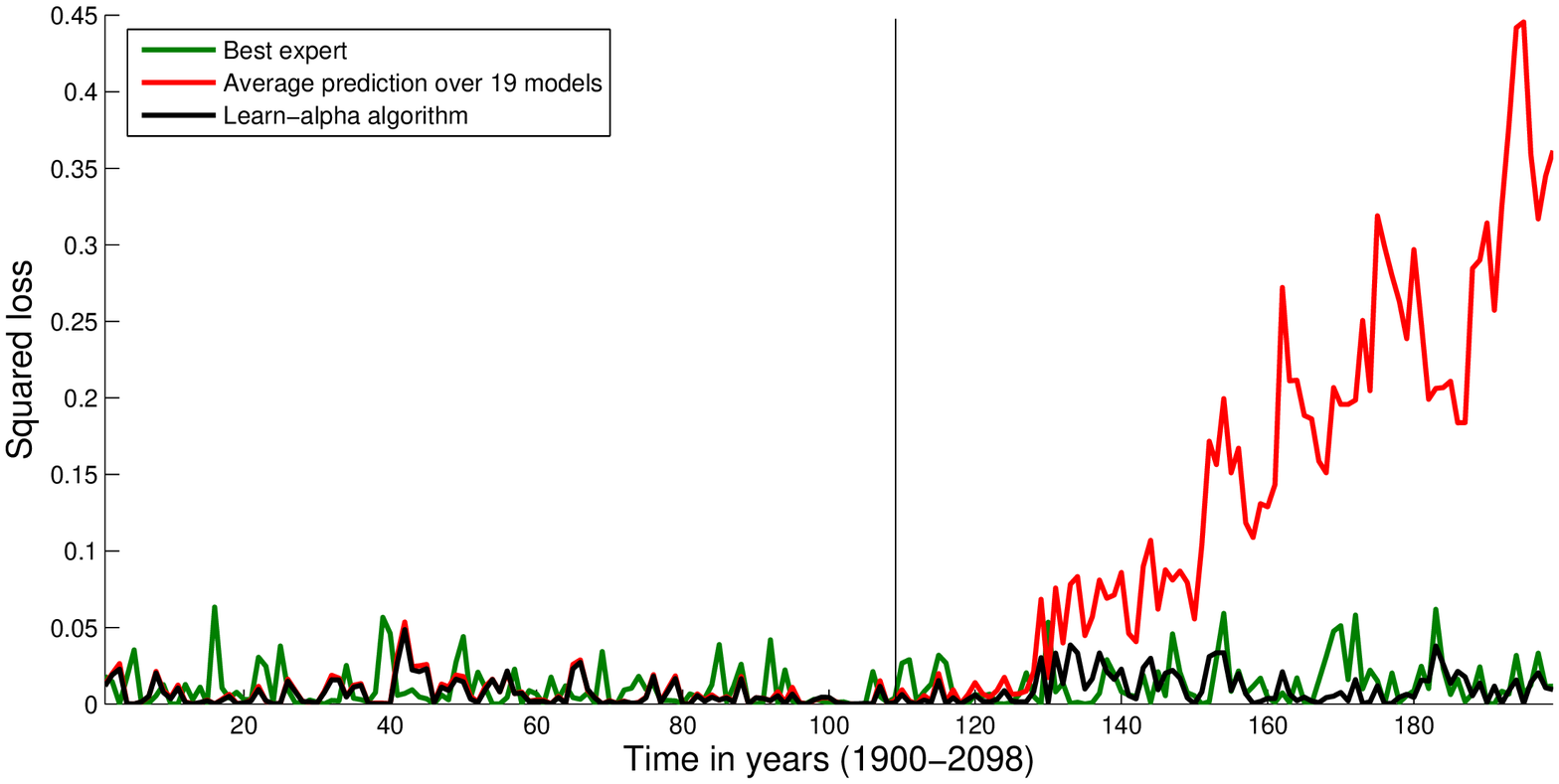}
\end{tabular}
\caption{Global Future Simulation 1: Tracking the predictions of one model
using the predictions of the remaining 19 as input, with no true temperature
observations.  Black vertical line separates past (hindcasts) from future
predictions.  Bottom plot zooms in on y-axis.  }\label{fig:futGiss}
\end{figure}

On historical global mean temperature data, our
online learning algorithm's average prediction loss nearly matches that of the
best performing climate model in hindsight. Moreover its performance
surpasses that of the average model prediction, which is the default practice
in climate science, the median prediction, and least squares linear regression.
We also experimented on climate model predictions through the year 2098.
Simulating labels with the predictions of any one climate model, we found
significantly improved performance using our online learning algorithm with
respect to the other climate models, and techniques (see \emph{e.g.}~Figure 1).  
To complement our global results, we also ran experiments on IPCC global
climate model temperature predictions for the specific geographic regions
of Africa, Europe, and North America.  On historical data, at both annual and
monthly time-scales, and in future simulations,  our algorithm typically
outperformed both the best climate model per region, and linear regression.
Notably, our algorithm consistently outperformed the average prediction
over models, the current benchmark.

\newpage
\setcounter{section}{0}
\setcounter{figure}{0}

\Title{Spectral Analysis Methods for Complex Source Mixtures} 
\bigskip\bigskip

\begin{raggedright}  
{\it Kevin H. Knuth \\
Departments of Physics and Informatics \\
University at Albany \\
Albany, New York, USA}
\bigskip\bigskip
\end{raggedright}

\section*{Abstract}

Spectral analysis in real problems must contend with the fact that there may 
be a large number of interesting sources some of which have known 
characteristics and others which have unknown characteristics.  In addition, 
one must also contend with the presence of uninteresting or background 
sources, again with potentially known and unknown characteristics.  In this 
talk I will discuss some of these challenges and describe some of the useful 
solutions we have developed, such as sampling methods to fit large numbers 
of sources and spline methods to fit unknown background signals.

\section*{Introduction}

The infrared spectrum of star-forming regions is dominated by emission from 
a class of benzene-based molecules known as Polycyclic Aromatic 
Hydrocarbons (PAHs).  The observed emission appears to arise from the 
combined emission of numerous PAH molecular species, both neutral and 
ionized, each with its unique spectrum.  Unraveling these variations is crucial 
to a deeper understanding of star-forming regions in the universe.  However, 
efforts to fit these data have been defeated by the complexity of the 
observed PAH spectra and the very large number of potential PAH emitters.
Linear superposition of the various PAH species accompanied by additional 
sources identifies this problem as a source separation problem. It is, 
however, of a formidable class of source separation problems given that 
different PAH sources are potentially in the hundreds, even thousands, and 
there is only one measured spectral signal for a given astrophysical site.  In 
collaboration with Duane Carbon
(NASA Advanced Supercomputing Center, NASA Ames),
we have focused on 
developing informed Bayesian source separation techniques \citep{knuth2005} 
to identify and characterize the contribution of a large number of PAH 
species to infrared spectra recorded from the Infrared Space Observatory 
(ISO).  To accomplish this we take advantage of a large database of over 
500 atomic and molecular PAH spectra in various states of ionization that 
has been constructed by the NASA Ames PAH team (Allamandola, 
Bauschlicher, Cami and Peeters).  To isolate the PAH spectra, much effort 
has gone into developing background estimation algorithms that model the 
spectral background so that it can be removed to reveal PAH, as well as 
atomic and ionic, emission lines.

\section*{The Spectrum Model}

Blind techniques are not always useful in complex situations like these where 
much is known about the physics of the source signal generation and 
propagation.  Higher-order models relying on physically-motivated 
parameterized functions are required, and by adopting such models, one can 
introduce more sophisticated likelihood and prior probabilities.  We call this 
approach Informed Source Separation \citep{knuth2007}.  In this problem, we 
have linear mixing of P PAH spectra, K Planck blackbodies, a mixture of G 
Gaussians to describe unknown sources and additive noise:

\begin{equation}\label{eqn1}
F(\lambda)=\sum_{p=1}^{P}c_{p}PAH_{p}(\lambda)+\sum_{k=1}^{K}A_{k}Planck(\lambda;T_{k})+\sum_{g=1}^{G}A_{g}N(\lambda;\bar{\lambda_{g}},\sigma_{g})+\phi(\lambda)
\end{equation}

where $PAH_{p}$ is a p-indexed PAH spectrum from the dictionary, $N$ is a 
Gaussian. The function Planck is

\begin{equation}\label{eqn2}
Planck(\lambda;T_{k})=\sqrt{\frac{\lambda_{max}}{\lambda}}\frac{exp(hc/\lambda_{max}kT) - 1}{exp(hc/\lambda kT) - 1}
\end{equation}

where $h$ is Planck's constant, $c$ is the speed of light, $k$ is Boltzmann's 
constant, $T$ is the temperature of the cloud, and $\lambda_{max}$
is the wavelength where the blackbody spectral energy peaks 
$\lambda_{max}=hc/4.965kT$.

\section*{Source Separation using Sampling Methods}

The sum over Planck blackbodies in the modeled spectrum (1) takes into 
account the fact that we are recording spectra from potentially several 
sources arranged along the line-of-sight.  Applying this model in conjunction 
with a nested sampling algorithm to data recorded from ISO of the Orion Bar 
we were able to obtain reasonable background fits, which often showed the 
presence of multiple blackbodies. The results indicate that there is one 
blackbody radiator at a temperature of 61.043 $\pm$ 0.004 K, and possibly a 
second (36.3\% chance), at a temperature around 18.8 K.
Despite these successes, this algorithm did not provide adequate results for 
background removal since the estimated background was not constrained to 
lie below the recorded spectrum.  Upon background subtraction, this led to 
unphysical negative spectral power.  This result encouraged us to develop 
an alternative background estimation algorithm.
Estimation of PAHs was demonstrated to be feasible in synthetic mixtures 
with low noise using sampling methods, such as Metropolis-Hastings Markov 
chain Monte Carlo (MCMC) and Nested Sampling.  Estimation using gradient 
climbing techniques, such as the Nelder-Mead simplex method, too often 
were trapped in local solutions.  In real data, PAH estimation was 
confounded by spectral background.

\section*{Background Removal Algorithm}

Our most advanced background removal algorithm was developed to avoid 
the problem of negative spectral power by employing a spline-based model 
coupled with a likelihood function that favors background models that lie 
below the recorded spectrum.  This is accomplished by using a likelihood 
function based on the Gaussian where the standard deviation on the 
negative side is 10 times smaller than on the positive side.
The algorithm is designed with the option to include a second derivative 
smoothing prior.  Users choose the number of spline knots and set their 
positions along the x-axis.  This provides the option of fitting a spectral 
feature or estimating a smooth background underlying it.
Our preliminary work shows that the background estimation algorithm works 
very well with both synthetic and real data \citep{nathan2010}.  The use of this 
algorithm illustrates that PAH estimates are extremely sensitive to 
background, and that PAH characterization is extremely difficult in cases 
where the background spectra are poorly understood.

\bigskip \bigskip
Kevin Knuth would like to acknowledge Duane Carbon, Joshua Choinsky, 
Deniz Gencaga, Haley Maunu, Brian Nathan and ManKit Tse for all of their 
hard work on this project.

\newpage
\setcounter{section}{0}
\setcounter{figure}{0}

\Title{Beyond Objects: Using Machines to Understand the Diffuse Universe}
\bigskip\bigskip
\begin{raggedright}  
{\it J. E. G. Peek\\
Department of Astronomy\\
Columbia University\\
New York, New York, USA}
\end{raggedright}  
\bigskip\bigskip


In this contribution I argue that our understanding of the universe has been shaped by an intrinsically ``object-oriented'' perspective, and that to better understand our diffuse universe we need to develop new ways of thinking and new algorithms to do this thinking for us. 

Envisioning our universe in the context of objects is natural both observationally and physically. When our ancestors looked up into the the starry sky, they noticed something very different from the daytime sky. The nighttime sky has specific objects, and we gave them names: Rigel, Procyon, Fomalhaut, Saturn, Venus, Mars. These objects were both very distinct from the blackness of space, but they were also persistent night to night. The same could not be said of the daytime sky, with its amorphous, drifting clouds, never to be seen again, with no particular identity. Clouds could sometimes be distinguished from the background sky, but often were a complex, interacting blend. From this point forward astronomy has been a science of objects. And we have been rewarded for this assumption: stars in space can be thought of very well as discrete things. They have huge density contrasts compared to the rest of space, and they are incredibly rare and compact. They rarely contact each other, and are typically easy to distinguish. The same can be said (to a lesser extent) of planets and galaxies, as well as all manner of astronomical objects. 

I argue, though, that we have gotten to a stage of understanding of our universe that we need to be able to better consider the diffuse universe. We now know that the material universe is largely made out of the very diffuse dark matter, which, while clumpy, is not well approximated as discrete objects. Even the baryonic matter is largely diffuse: of the 4\% of the mass-energy budget of the universe devoted to baryons, 3.5\% is diffuse hot gas permeating the universe, and collecting around groups of galaxies. Besides the simple accounting argument, it is important to realize that the interests of astronomers are now oriented more and more toward origins: origins of planets, origins of stars, origins of galaxies. This is manifest in the fact that NASA devotes a plurality of its astrophysics budget to the ``cosmic origins'' program. And what do we mean by origins? The entire history of anything in the universe can be roughly summed up as ``it started diffuse and then, under the force of gravity, it became more dense''. If we are serious about understanding the origins of things in the universe, we must do better at understanding not just the objects, but the diffuse material whence they came.

We have, as investigators of the universe, enlisted machines to do a lot of our understanding for us. And, as machines inherit our intuition through the codes and algorithms we write, we have given them a keen sense of objects. A modern and powerful example is the Sloan Digital Sky Survey \citep[SDSS;][]{york00}. SDSS makes huge maps of the sky with very high fidelity, but these maps are rarely used for anything beyond wall decor. The real power of the SDSS experiment depends on the photometric pipeline \citep{Lupton01}, which interprets that sky into tens of millions of objects, each with precise photometric information. With these lists in hand we can better take a census of the stars and galaxies in our universe. It is sometimes interesting to understand the limits of these methodologies; the photo pipeline can find distant galaxies easily, but large, nearby galaxies are a challenge, as the photo pipeline cannot easily interpret these huge diaphanous shapes \citep[][Fig 1]{West10}. The Virtual Astronomical Observatory \citep[VAO; e.g.][]{Hanisch10} is another example of a collection of algorithms that enables our object-oriented mindset. VAO has developed a huge set of tools that allow astronomers to collect a vast array of information from different sources, and combine them elegantly together. These tools, however, almost always use the ``object" as the smallest element of information, and are much less useful in interpreting the diffuse universe. Finally, astrometry.net is an example of how cutting edge algorithms combined with excellent data can yield new tools for interpreting astronomical data \citep{Lang10}. By accessing giant catalogs of objects, the software can, in seconds, give precise astrometric information about any image containing stars. Again, we leverage our object-oriented understanding, both both psychologically and computationally, to decode our data.

\begin{figure}[htbp]
\begin{center}
\includegraphics[scale=1.5]{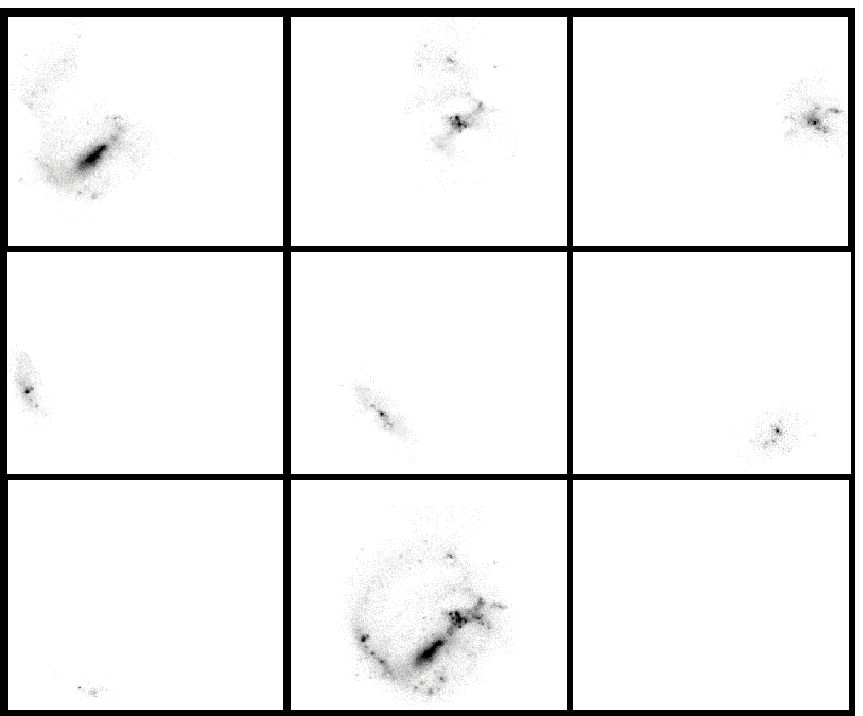}
\caption{\emph{r}-band atlas images for HIPEQ1124+03. A single galaxy has been divided into 7 sub-images by the SDSS photometric pipeline, which considered them individual objects. The original galaxy is shown in the lower-middle panel. Reprinted with permission from \citet{West10}.}
\label{photofail}
\end{center}
\end{figure}

As a case study, we examine at a truly object-less data space: the Galactic neutral hydrogen (\hia) interstellar medium (ISM). Through the 21-cm hyperfine transition of \hia, we can study the neutral ISM of our Galaxy and others both angularly and in the velocity domain \citep[e.g.][]{KH88}. \hi images of other galaxies, while sometimes diffuse, do typically have clear edges. In our own Galaxy we are afforded no such luxury. The Galactic \hi ISM is sky-filling, and can represent gas on a huge range of distances and physical conditions. As our technology increases, we are able to build larger and larger, and more and more detailed images of the \hi ISM. What we see in these multi-spectral images is an incredible cacophony of shapes and structures, overlapping, intermingling, with a variety of size, shape, and intensity that cannot be easily described. Indeed, it is this lack of language that is at the crux of the problem. These data are affected by a huge number of processes; the accretion of material onto the Galaxy \citep[e.g.][]{Begum10}, the impact of shockwaves and explosions \citep[e.g.][]{Heiles79}, the formation of stars \citep[e.g.][]{Kim98}, the effect of magnetization \citep[e.g.][]{McC-G06a}. And yet, we have very few tools that capture this information.

\begin{figure*}
\begin{center}
\includegraphics[scale=.42, angle=0]{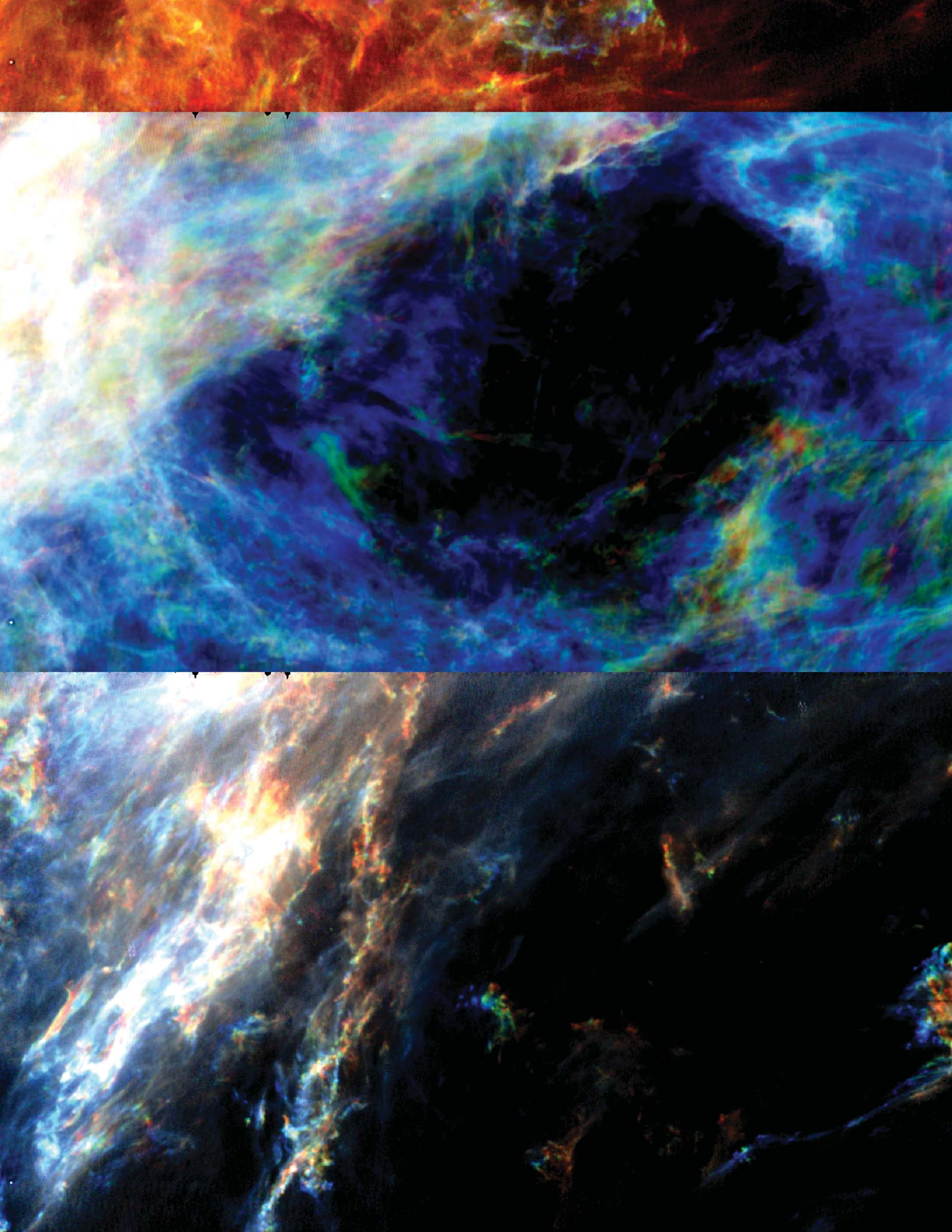}
\caption{A typical region of the Galactic \hi sky, 40$^\circ \times18^\circ$ $10^\prime$ in size. The top panel represents, -41.6, -39.4, -37.2 \kms in red, green, and blue, respectively. The middle panel represents -4.0, -1.8, and 0.4 \kmsa, while the bottom panel represents 15.8, 18.7, 21.7 \kmsa. Reprinted with permission from \citep{Peek11}. \label{slices}}
\end{center}
\end{figure*}

As yet, there are two ``flavors'' of mechanisms we as a community have used to
try to interpret this kind of diffuse data. The first is the observer's method.
In the observer's method the data cubes are inspected by eye, and visually
interesting shapes have been picked out  \citep[e.g.][]{FLM-G10}. These shapes
are then cataloged and described, usually qualitatively and without statistical
rigor. The problems with these methods are self-evident: impossible statistics,
unquantifiable biases, and an inability to compare to physical models. The
second method is the theorist's method. In the theorist's method, some equation
is applied to the data set wholesale, and a number comes out
\citep[e.g.][]{Chepurnov10}. This method is powerful in that it can be compared
directly to simulation, but typically cannot interpret any shape information at
all. Given that the ISM is not a homogeneous, isotropic system, and various
physical effects may influence the gas in different directions or at different
velocities, this method seems a poor match for the data. It also cuts out any
intuition as to what data may be carrying the most interesting information.

\begin{figure}[htbp]
\begin{center}
\includegraphics[scale=1.50]{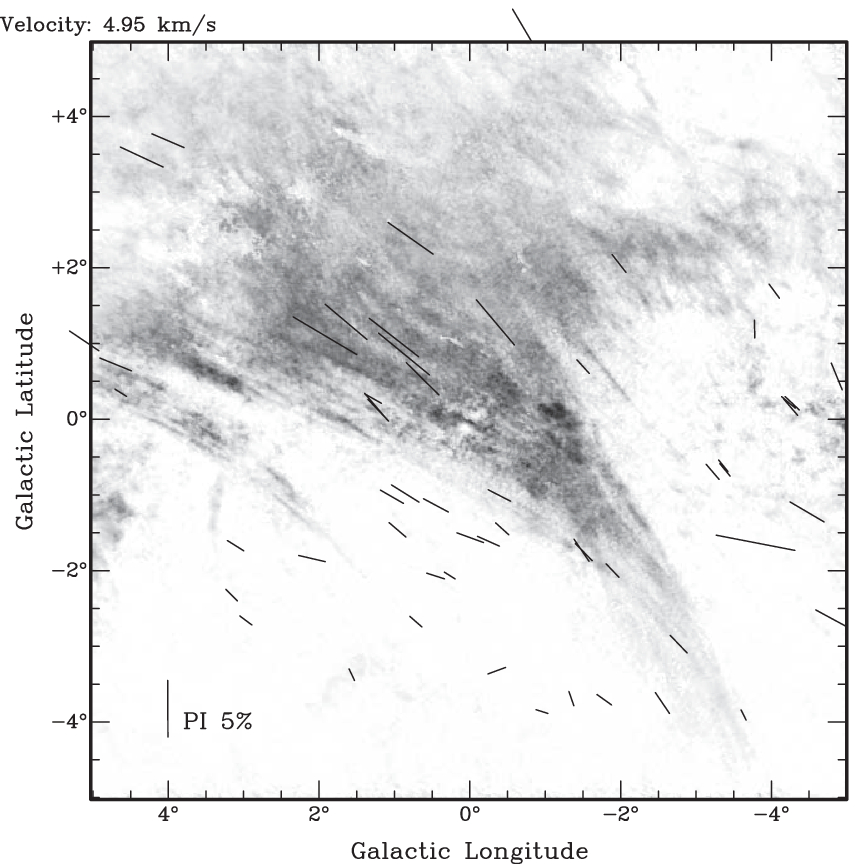}
\caption{An \hi image of the Riegel-Crutcher cloud from \citet{McC-G06a} at
4.95 \kms $v_{\rm LSR}$. The polarization vectors from background starlight
indicate that the structure of the ISM reflects the structure of the intrinsic
magnetization. Reprinted with permission from \citet{McC-G06a}.}
\label{rccloud}
\end{center}
\end{figure}

We are in the process of developing a ``third way'', which I will explain in two examples. Of the two projects, our more completed one is a search for compact, low-velocity clouds in the Galaxy  \citep[e.g.][]{Saul11}. These clouds are inherently interesting as they likely probe the surface of the Galaxy as it interacts with the Galactic halo, a very active area of astronomical research. To do this our group, led by Destry Saul, wrote a wavelet-style code to search through the data cubes for isolated clouds that matched our search criteria. These clouds once found could then be ``objectified'', quantified and studied as a population. In some sense, through this objectification, we are trying to shoehorn an intrinsically diffuse problem into the object-oriented style thinking we are trying to escape. This gives us the advantage that we can use well known tools for analysis (e.g. scatter plots), but we give up a perhaps deeper understanding of these structures from considering them in their context. The harder, and far less developed, project is to try to understand the meaning of very straight and narrow diffuse structures in the HI ISM at very low velocity. The HI ISM is suffused with ``blobby filaments'', but these particular structures seem to stand out, looking like a handful of dry fettuccine dropped on the kitchen floor. We know that these kinds of structures can give us insight into the physics of the ISM: in denser environments it has been shown that more discrete versions of these features are qualitatively correlated with dust polarizations and the magnetic underpinning of the ISM \citep{McC-G06a}. We would like to investigate these features more quantitatively, but we have not developed mechanisms to answer even the simplest questions. In a given direction how much of this feature is there? In which way is it pointing? What are its qualities? Does there exist a continuum of these features, or are they truly discrete? The ``object-oriented'' astronomer mindset is not equipped to address these sophisticated questions.

We are just beginning to investigate machine vision techniques for understanding these unexplored data spaces. Machine vision technologies are being developed to better parse our very confusing visual world using computers, such as in the context of object identification and the 3D reconstruction of 2D images \citep{Sonka08}. Up until now, most astronomical machine vision problems have been embarrassingly easy; points in space are relatively simple to parse for machines. Perhaps the diffuse universe will be a new challenge for computer vision specialists and be a focal point for communication between the two fields. Machine learning methods, and human-aided data interpretation on large scales may also prove crucial to cracking these complex problems. How exactly we employ these new technologies in parsing our diffuse universe is very much up to us.


\newpage
\setcounter{section}{0}
\setcounter{figure}{0}

\Title{Viewpoints: A high-performance high-dimensional exploratory data
analysis tool}
\bigskip\bigskip
\begin{raggedright}  

{\it Michael Way\\
NASA/Goddard Institute for Space Studies\\
2880 Broadway\\
New York, New York, USA}
\bigskip

{\it Creon Levit \& Paul Gazis\\
NASA/Ames Research Center\\
Moffett Field, California, USA}

\bigskip\bigskip
\end{raggedright}  


Viewpoints \citep{Gazis2010} is a high-performance visualization and analysis
tool for large,
complex, multidimensional data sets. It allows interactive exploration of data
in 100 or more dimensions with sample counts, or the number of points, exceeding
$10^{6}$ (up to $10^{8}$ depending on available RAM). Viewpoints was originally
created for use with the extremely large data sets produced by current
and future NASA
space science missions, but it has been used for a wide variety of diverse
applications ranging from aeronautical engineering, quantum chemistry, and
computational fluid dynamics to virology, computational finance, and aviation
safety. One of it's main features is the ability to look at the correlation
of variables in multivariate data streams (see Figure 1).

Viewpoints can be considered a kind of ``mini" version of the NASA Ames
Hyperwall \citep{Sandstrom2003} which has been used for examining multi-variate
data of much larger sizes (see Figure 2). Viewpoints has been used extensively
as a pre-processor to the Hyperwall in that one can look at sub-selections
of the full dataset (if the full data set cannot be run) prior to viewing
it with the Hyperwall (which is a highly leveraged resource). Currently
viewpoints runs on Mac OS, Windows and Linux platforms, and only requires
a moderately new (less than 6 years old) graphics card supporting OpenGL.

More information can be found here:\newline
http://astrophysics.arc.nasa.gov/viewpoints\newline
You can download the software from here:\newline
http://www.assembla.com/wiki/show/viewpoints/downloads\newline

\begin{figure}[!htb]
\includegraphics[scale=0.4]{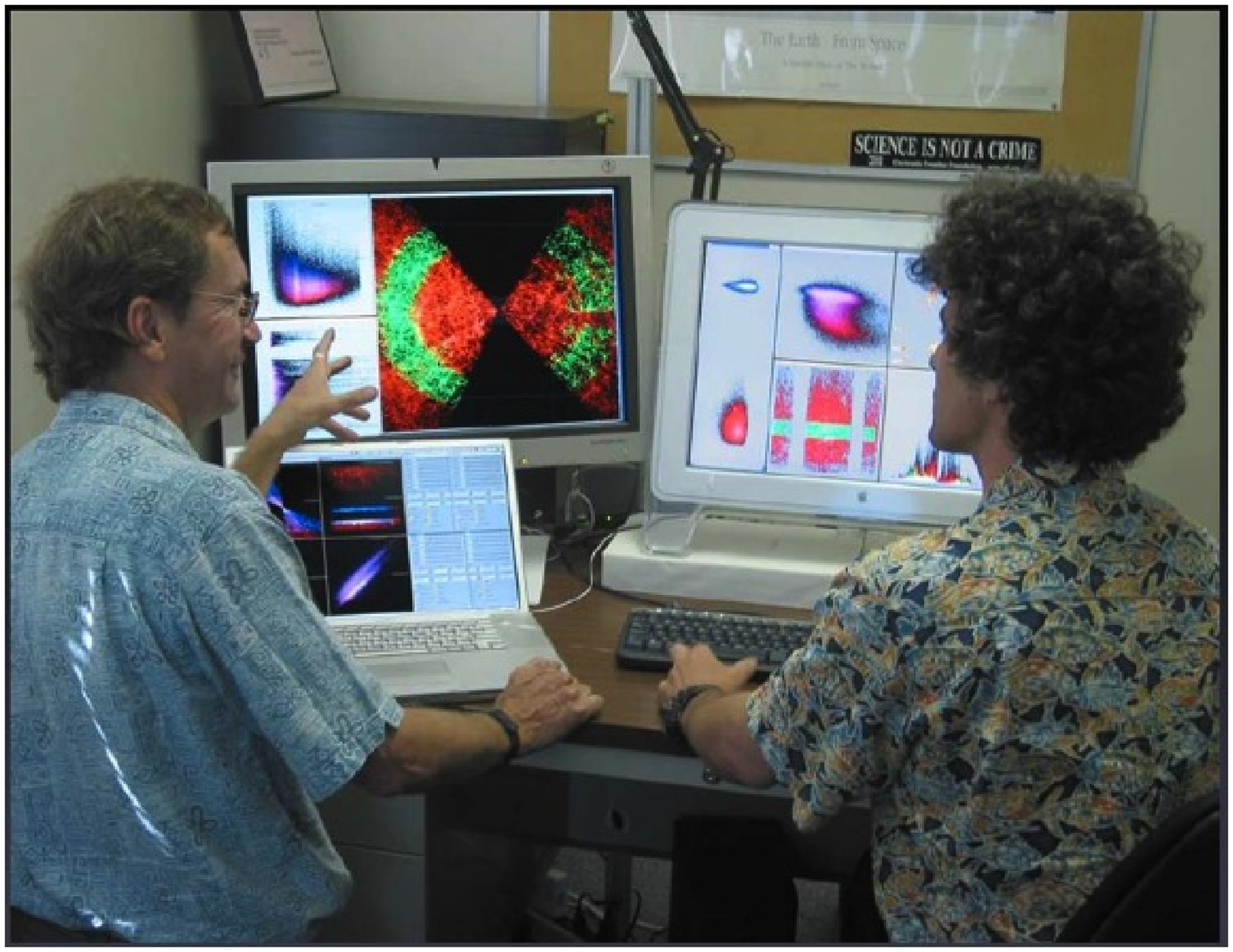}
\caption{Viewpoints as a collaboration tool: Here one workstation with multiple
screens is looking at the same multi-variate data on a laptop. Screen layout and
setup can be saved to an xml file which allows one to retrace previous
investigations.}
\label{fig1}
\end{figure}

\begin{figure}[!htb]
\includegraphics[scale=0.29]{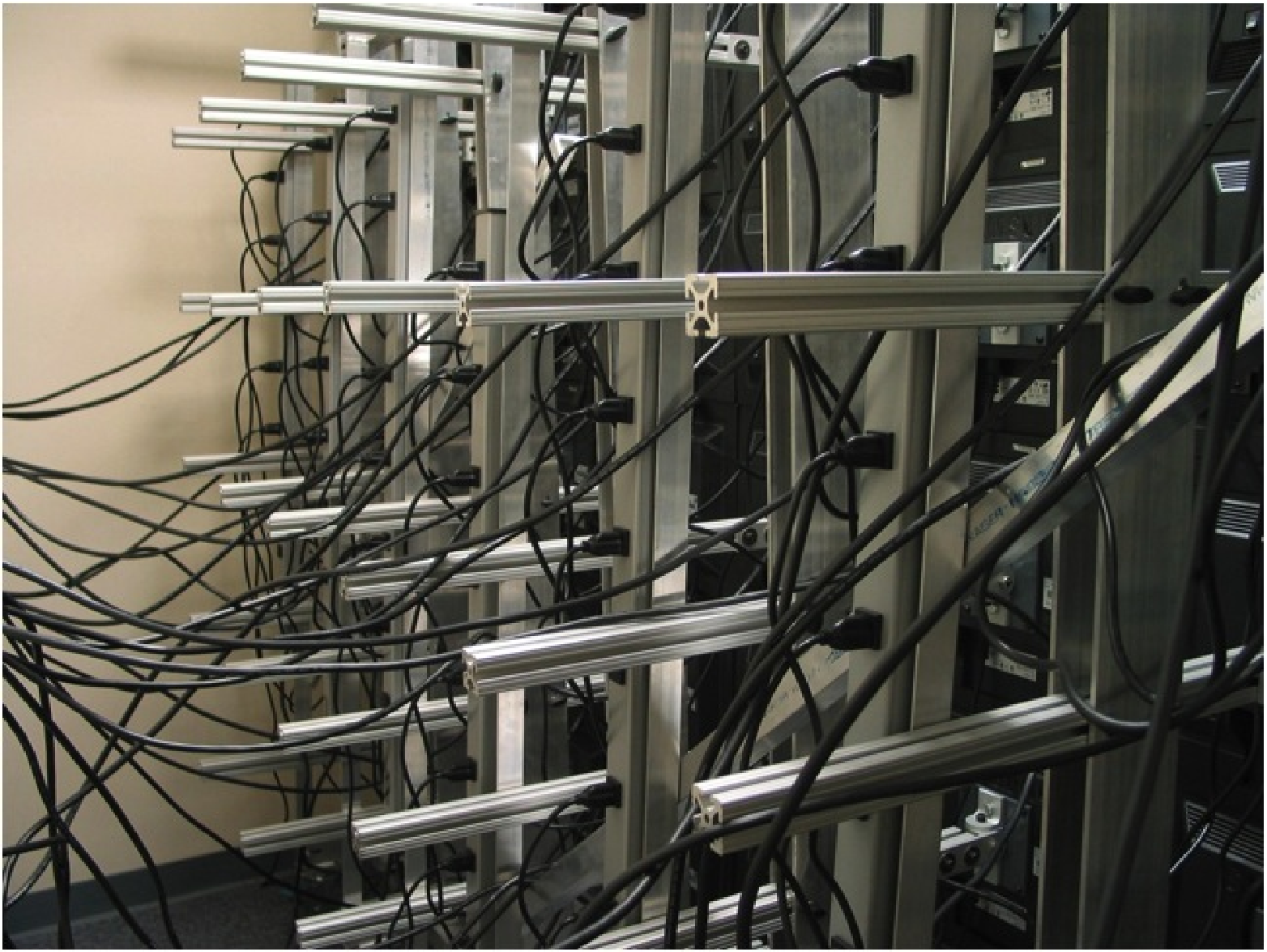}
\includegraphics[scale=0.284]{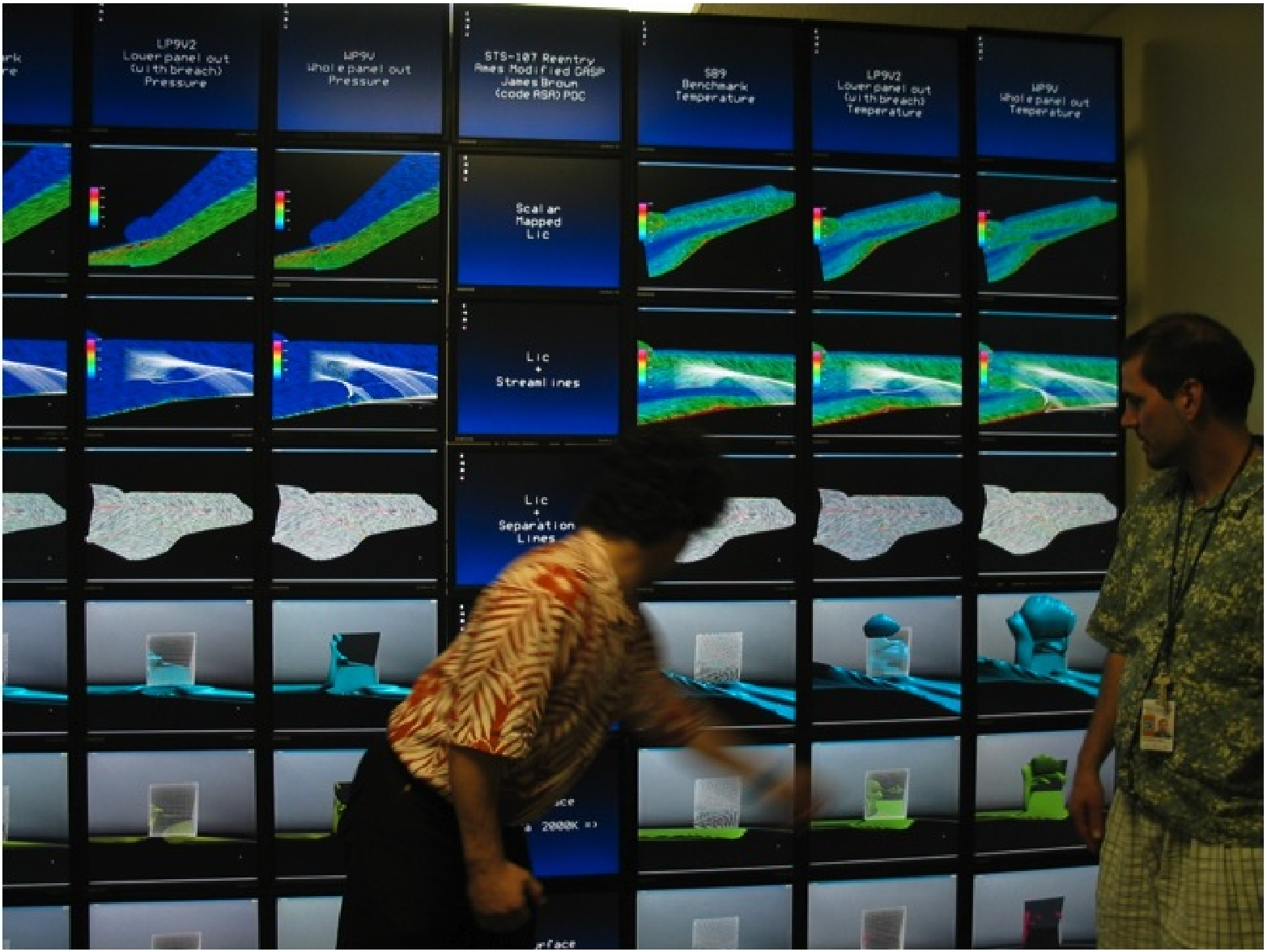}
\caption{Left: The back of the original (7$\times$7 display) hyperwall
at NASA/Ames. Right: The front of the hyperwall. One can see the obvious
similarities between the Hyperwall and viewpoints.}
\label{fig2}
\end{figure}

\newpage
\setcounter{section}{0}
\setcounter{figure}{0}

\Title{Clustering Approach for Partitioning Directional Data in Earth and
Space Sciences}

\bigskip\bigskip
\begin{raggedright}
{\it C. D. Klose\\
Think GeoHazards\\
New York, New York, USA}

\bigskip

{\it K. Obermayer\\
Electrical Engineering and Computer Science\\
Technical University of Berlin\\
Berlin, Germany}
\bigskip\bigskip
\end{raggedright}

\section*{Abstract}
A simple clustering approach, based on vector quantization (VQ) is
presented for partitioning directional data in Earth and Space
Sciences. Directional data are grouped into a certain number of
disjoint isotropic clusters, and at the same time the average
direction is calculated for each group. The algorithm is fast, and
thus can be easily utilized for large data sets. It shows good
clustering results compared to other benchmark counting methods for
directional data. No heuristics is being used, because the grouping
of data points, the binary assignment of new data points to clusters,
and the calculation of the average cluster values are based on the
same cost function.

\bigskip
\noindent {\it Keywords: } clustering, directional data, discontinuities,
fracture grouping
\bigskip

\section*{Introduction}
\label{sec:intro}

Clustering problems of directional 
data are fundamental problems in earth and space sciences.
Several methods have been proposed to help to find groups within 
directional data. 
Here, we give short overview on existing clustering methods of directional data and outline 
a new clustering method which is based on vector quantization \citep{Gray84}. The new method
improves on several issues of clustering directional data and is published by \cite{Klose04}. 

Counting methods for visually partitioning the orientation data in stereographic plots were 
introduced by \cite{Schmidt25}. 
\cite{ShanleyMathab76} and \cite{Wall78} developed counting 
techniques to identify clusters of orientation data. The parameters of Shanley \& Mahtab's 
counting method have to be optimized by minimizing an objective function.   
Wallbrecher's method is optimized by comparing the clustering result with a given 
probability distribution on the sphere in order to obtain good partitioning results. 
However, counting methods depend on the density of data points and their 
results are prone to sampling bias (e.g., 1-D or 2-D sampling to describe a 3-D space). 
Counting methods are time-consuming, can lead to incorrect results for clusters with 
small dip angles, and can lead to solutions which an expert would rate sub-optimal.
\cite{Pecher89} developed a supervised method for grouping of directional data 
distributions. A contour density plot is calculated and an observer picks initial values 
for the average dip directions and dip angles of one to a maximum of seven clusters. 
The method has a conceptual disadvantage. It uses two different distance
measures; one measure for the assignment of data points to clusters and another 
measure defined by the orientation matrix to calculate the refined values for dip 
direction and dip angle. Thus, average values and cluster assignments are not determined 
in a self-consistent way. 

\cite{Dershowitz96} developed a partitioning method that is based 
on an iterative, stochastic reassignment of orientation vectors
to clusters. Probability assignments are calculated using selected probability
distributions on the sphere, which are centered on the average orientation
vector that characterizes the cluster. The average orientation vector is
then re-estimated using principal component analysis (PCA) of the orientation
matrices. Probability distributions on the sphere were developed by several 
authors and are summarized in \cite{Fisher87}. 

\cite{Hammah98} described a related approach based on 
fuzzy sets and on a similarity measure $d^2(\vec{x}, \vec{w}) = 1 - (\vec{x}^T \vec{w})^2$, 
where $\vec{x}$ is the orientation vector of a data point and $\vec{w}$ is the 
average orientation vector of the cluster. This measure is normally 
used for the analysis of orientation data \citep{Anderberg73,Fisher87}. 

\section*{Directional Data}

Dip direction $\alpha$ and the dip angle $\theta$ of linear or planar structures 
are measured in degrees ($^{\circ}$), where $0^{\circ} \leq
\alpha \leq 360^{\circ}$ and $0^{\circ} \leq \theta \leq 90^{\circ}$.
By convention, linear structures and normal vectors of planar structures, 
{\it pole vectors} $\vec{\Theta} = (\alpha, \theta)^T$, point towards the lower 
hemisphere of the unit sphere (Figure 1).
The orientation $\vec{\Theta}^A = (\alpha^A,
\theta^A)^T$ of a pole vector $A$ can be described by Cartesian
coordinates $\vec{x}^A = (x_1, x_2, x_3)^T$ (Figure 1), where\\
\begin{equation}
\begin{split}
&x_1 \;=\; cos(\alpha) \, cos(\theta) \hspace{1cm} \text{North direction}\\
&x_2 \;=\; sin(\alpha) \, cos(\theta) \hspace{1cm} \text{East direction}\\
&x_3 \;=\; sin(\theta) \hspace{2.3cm} \text{downward}.
\label{eq:dir_cosines}
\end{split}
\end{equation}

The projection $A'$ of the endpoint $A$ of all given pole vectors onto the 
$x_1$-$x_2$ plane is called a stereographic plot (Figure 1)
and is commonly used for visualisation purposes.

\begin{center}
\begin{minipage}[t]{16cm}
  A)\includegraphics[width=6.5cm]{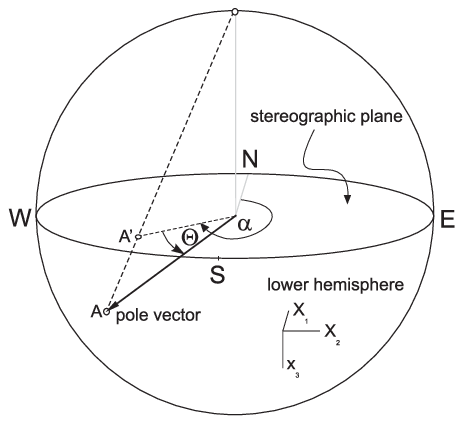}
  B)\includegraphics[width=5.5cm]{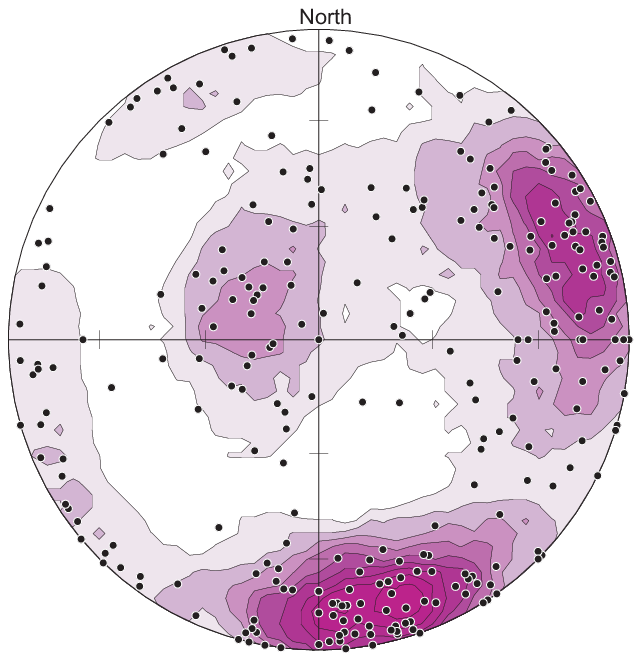}
\end{minipage}
\parbox[t]{16cm}{\footnotesize{\textbf{Figure 1:}
A) Construction of a stereographic plot.\newline
B) Stereographic plot with kernel density distribution.}}
\end{center}

\section*{The Clustering Method}

Given are a set of $N$ pole vectors $\vec{x}_k$, $k=1,\ldots,N,$
(eq.\ \ref{eq:dir_cosines}). The vectors correspond to $N$ noisy 
measurements taken from $M$ orientation discontinuities whose spatial 
orientations are described by their
(yet unknown) average pole vectors $\vec{w}_l$, $l=1,\ldots,M$. For
every partition $l$ of the orientation data, there exists one average
pole vector $\vec{w}_l$. The dissimilarity between a data point
$\vec{x}_k$ and an average pole vector $\vec{w}_l$ is denoted by
$d(\vec{x}_k,\vec{w}_l)$.

We now describe the assignment of pole vectors $\vec{x}_k$ to a
partition by the binary assignment variables
\begin{equation}
m_{lk} \;=\; \left\{\begin{array}{ll}
1, & \textrm{if data point $k$ belongs to cluster $l$}\\
0, & \textrm{otherwise.}
\end{array} \right .\label{eq:assign_variables}
\end{equation}
One data point $\vec{x}_k$ belongs to only one orientation discontinuity $\vec{w}_l$. 
Here, the arc-length between the pole vectors on the unit sphere is proposed as 
the distance measure, i.e.
\begin{equation}
d(\vec{x},\vec{w}) \;=\;\arccos\, (\,|\,\vec{x}^T \vec{w}\,|\,),
\label{eq:distance}  
\end{equation}
where $|.|$ denotes the absolute value. 

The average dissimilarity between the data points and the pole vectors of the directional data
they belong to is given by
\begin{equation}
E \;=\; \frac{1}{N} \sum_{k=1}^N\sum_{l=1}^M m_{lk}\, d(\vec{x}_k,\vec{w}_l),
\label{eq:cost1}
\end{equation}
from which we calculate the optimal partition by minimizing the cost
function $E$, i.e.\
\begin{equation}
E \;\stackrel{!}{=}\; \min_{\{m_{lk}\},\{\vec{w}_l\}}.
\label{eq:cost_min}
\end{equation}

Minimization is performed iteratively in two steps. In the first step, the
cost function $E$ is minimized with respect to the assignment variables $\{m_{lk}\}$
using
\begin{equation}
m_{lk} \;=\; \left\{\begin{array}{ll}
1, & \mbox{\rm if} \;\;\; l \,=\, \arg \min_q \, d(\vec{x}_k,\vec{w}_q)\\
0, & \mbox{\rm else.}
\end{array} \right .\label{eq:assignment}
\end{equation}
In the second step, cost $E$ is minimized with respect to the angles $\vec{\Theta_l}
= (\alpha_l, \theta_l)^T$ which describe the average pole vectors $\vec{w}_l$
(see eq.\ (\ref{eq:dir_cosines})).
This is done by evaluating the expression
\begin{equation}
\frac{\partial E}{\partial \vec{\Theta}_l} \;=\; \vec{0},
\label{eq:weight_update}
\end{equation}
where $\vec{0}$ is a zero vector with respect to $\vec{\Theta}_1 = (\alpha_l, \theta_l)^T$.
This iterative procedure is called batch learning and converges to a minimum
of the cost, because $E$ can never increase and is bounded from below. In
most cases, however, a stochastic learning procedure called on-line learning
is used which is more robust:
\begin{center}
\parbox[t]{10truecm}{
BEGIN Loop\\
\null \hskip 0.5truecm Select a data point $\vec{x}_k$.\\
\null \hskip 0.5truecm Assign data point $\vec{x}_k$ to cluster $l$ by:
\begin{equation}
l \;=\; \arg \min_q d(\vec{x}_k,\vec{w}_q)
\label{eq:online_assign}
\end{equation}
\hskip 0.5truecm Change the average pole vector of this cluster by:
\begin{equation}
\Delta \vec{\Theta}_l \;=\; - \gamma \frac{\partial d(\vec{x}_k,
\vec{w}_l(\vec{\Theta}_l))}{\partial \vec{\Theta}_l}
\label{eq:online_update}
\end{equation}
END Loop
}
\end{center}
The learning rate $\gamma$ should decrease with iteration number $t$, such
that the conditions \citep{RobbinsMonro51,Fukunaga90}
\begin{equation}
\sum_{t=1}^\infty \gamma(t)\;=\,\infty,\quad \textrm{and}
\quad \sum_{t=1}^\infty \gamma^2(t)\;<\;\infty
\end{equation}
are fulfilled.

\section*{Results}

The clustering algorithm using the arc-length as distance measure is derived and applied in 
Klose et al.~ (2004) and online available as a Java app (http://www.cdklose.com). 
First, the new clustering algorithm is applied to an artificial data 
set where orientation and distribution of pole vectors are statistically defined in advance. 
Second, the algorithm is applied to a real-world example given by \cite{ShanleyMathab76} 
(see Figure 1). Results are compared to existing counting and clustering methods, 
as described above. 

\begin{table}[htb]
\begin{center}
\begin{tabular}[t]{|c|c|}\hline
Input& Output \\\hline
& \\
\includegraphics[width=6cm]{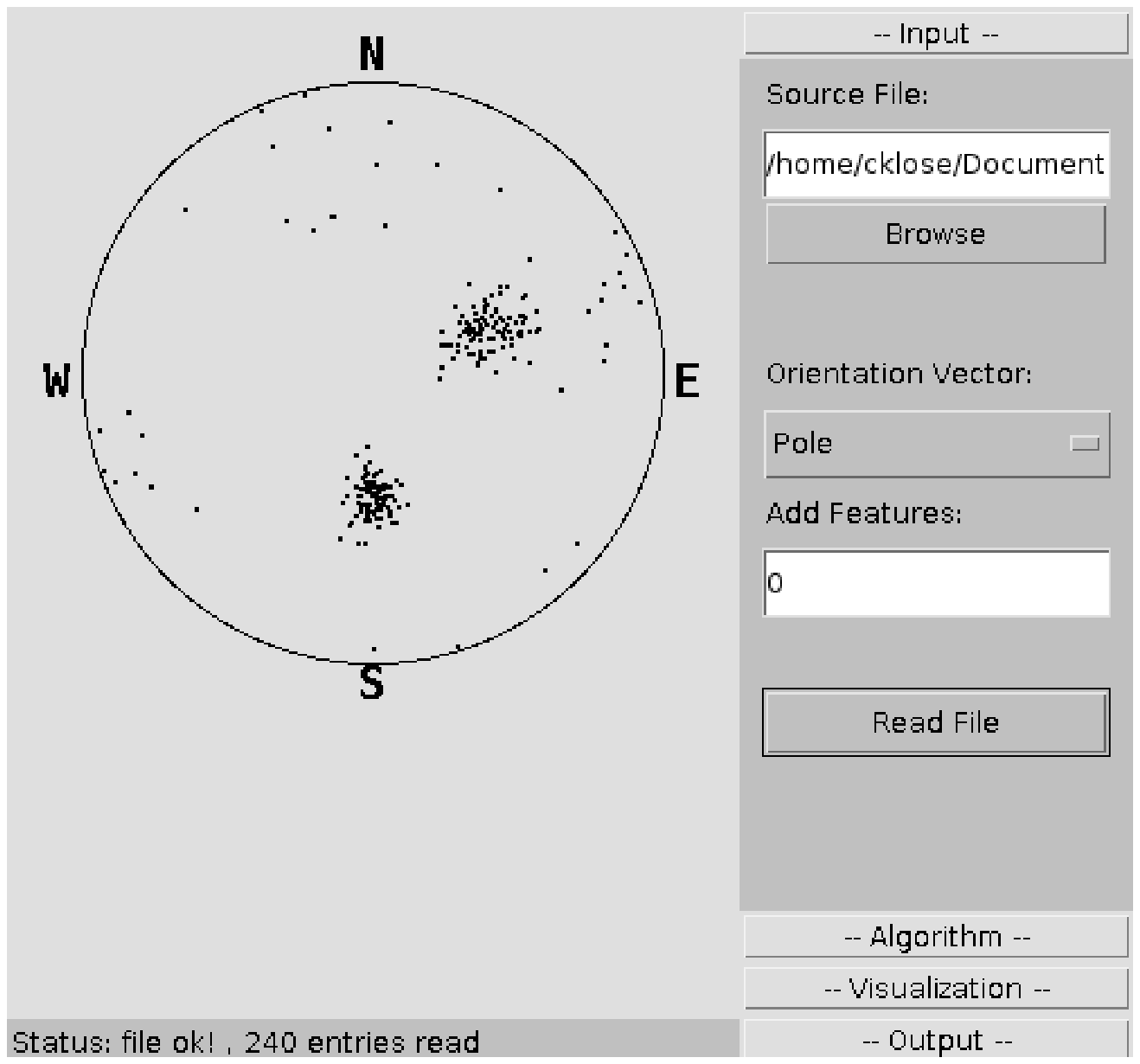} &\includegraphics[width=6cm]{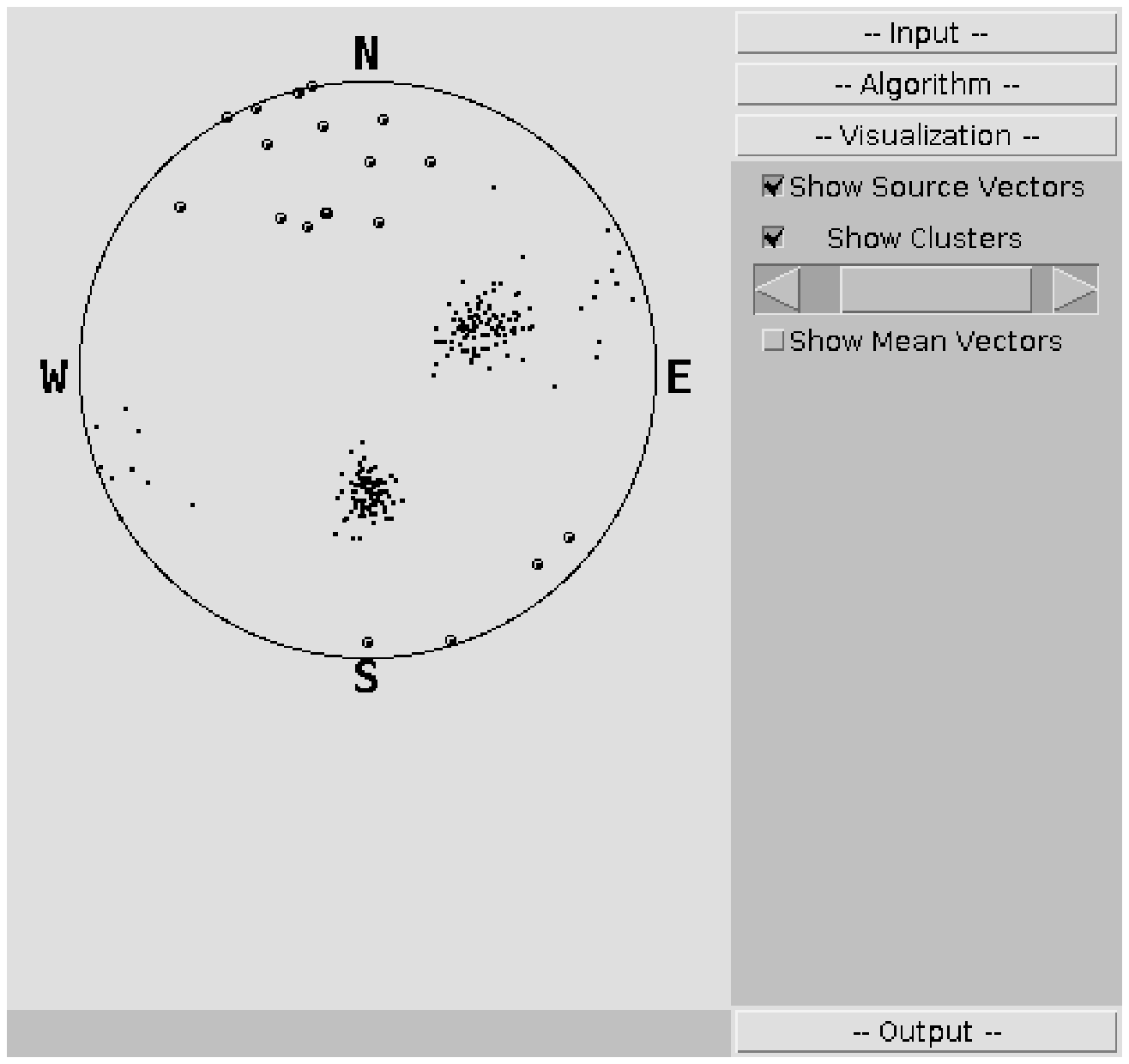}\\
& \\\hline
\end{tabular}
\end{center}
Figure 2: Snapshots of the Java app of clustering algorithm available at
http://www.cdklose.com
\end{table}

\newpage
\setcounter{section}{0}
\setcounter{figure}{0}

\Title{Planetary Detection: The Kepler Mission}

\begin{raggedright}  
{\it Jon Jenkins\\
NASA/Ames Research Center\\
Moffett Field, California, USA}
\bigskip\bigskip
\end{raggedright}

The Kepler telescope was launched into orbit in March 2009 to determine the
frequency of Earth-sized planets transiting their Sun-like host stars in the
habitable zone -- that range of orbital distances for which liquid water would
pool on the surface of a terrestrial planet such as Earth or Mars. This daunting
task demands an instrument capable of measuring the light output from each
of over 150,000 stars over a 115 square degree field of view simultaneously at
an unprecedented photometric precision of 20 parts per million (ppm) on
6.5-hour intervals. Kepler is opening up a new vista in astronomy and
astrophysics
and is operating in a new regime where the instrumental signatures compete
with the miniscule signatures of terrestrial planets transiting their host
stars.  The dynamic range of the intrinsic stellar variability observed in the
light curves
is breathtaking: RR Lyrae stars explosively oscillate with periods of
approximately 0.5 days, doubling their brightness over a few hours. Some flare
stars double their brightness on much shorter time scales at unpredictable
intervals. At the same time, some stars exhibit quasi-coherent oscillations with
amplitudes of 50 ppm that can be seen by eye in the raw flux time series. The
richness of Kepler's data lies in the huge dynamic range for the variations in
intensity $>$10$^{4}$ and the large dynamic range of time scales probed by
the data, from a few minutes to weeks, months, and ultimately, to years.

Kepler is an audacious mission that places rigorous demands on the science
pipeline used to process the ever-accumulating, large amount of data and to
identify and characterize the minute planetary signatures hiding in the data
haystack. We give an overview of the Science pipeline that reduces the pixel
data to obtain flux time series and detect and characterize planetary transit
signatures. In particular, we detail the adaptive, wavelet-based transit
detector that performs the automated search through each light curve for transit
signatures of Earth-sized planets. We describe a Bayesian Maximum A
Posteriori (MAP) estimation approach under development to improve our
ability to identify and remove instrumental signatures from the light curves
that minimizes any distortion of astrophysical signals in the data and prevents
the introduction of additional noise that may mask small, transit features, as
indicated in the Figure 1. This approach leverages the availability of thousands
of stellar targets on each CCD detector in order to construct an implicit
forward model for the systematic error terms identified in the data as a whole.
The
Kepler Mission will not be the last spaceborne astrophysics mission to scan the
heavens for planetary abodes. Several transit survey missions have been
proposed to NASA and to ESA and some are under development. Clearly, these
future missions can benefit from the lessons learned by Kepler and will face
many of the same challenges that in some cases will be more difficult to solve
given the significantly larger volume of data to be collected on a far greater
number of stars than Kepler has had to deal with. Given the intense interest in
exoplanets by the public and by the astronomical community, the future for
exoplanet science appears to just be dawning with the initial success of the
Kepler Mission.

\begin{figure}
\includegraphics[scale=0.9]{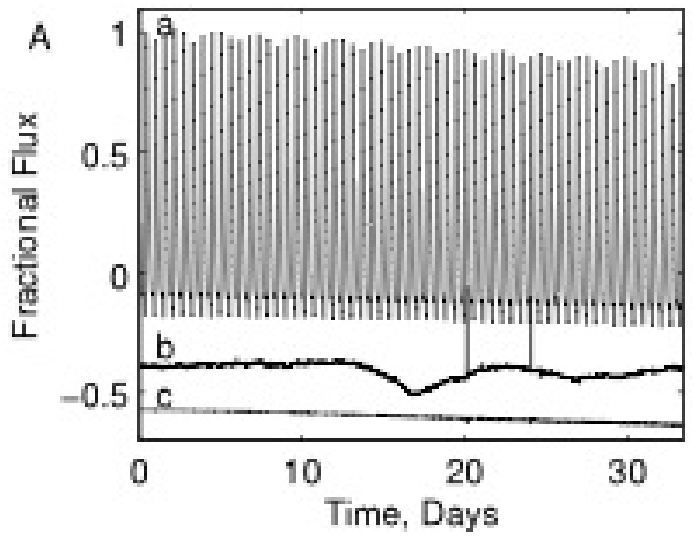}
\includegraphics[scale=0.9]{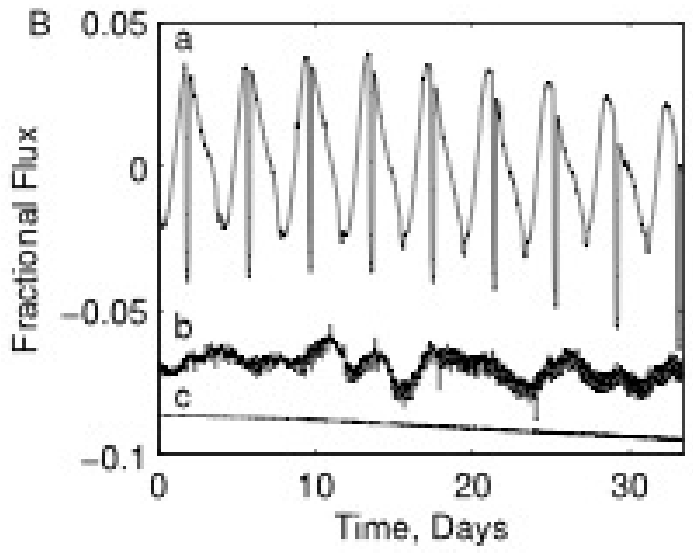}
\caption{\footnotesize
The light curves (a) for two stars on channel 2.1, along with (b) an LS fit to
instrumental components extracted from the light curves and (c) a Bayesian
Maximum A Posteriori (MAP) fit to the same instrumental components. Curves
(b) and (c) have been offset from 0 for clarity. Panel A shows the results for
an RR Lyrae star while panel B shows them for an eclipsing binary. Both light
curves are dominated by intrinsic stellar variability rather than by
instrumental
signatures. The RR Lyrae doubles its brightness every 0.5 days, while the
eclipsing binary exhibits spot variations that change slowly over time. The MAP
fits do not corrupt the data with short term variations in the poorly matched
instrumental signatures used in the fit, unlike the least squares fit.
}
\label{jenkins-figs}
\end{figure}

\newpage
\setcounter{section}{0}
\setcounter{figure}{0}

\Title{Understanding the possible influence of the solar activity on the
terrestrial climate: a times series analysis approach}
\bigskip\bigskip
\begin{raggedright}  

{\it Elizabeth Mart\'{i}nez-G\'{o}mez \\
Center for Astrostatistics\\
326 Thomas Building\\
The Pennsylvania State University\\
University Park, PA, USA}
\bigskip

{\it V\'{i}ctor M. Guerrero \\
Departmento de Estad\'{i}stica\\
Instituto Tecnol\'{o}gico Aut\'{o}nomo de M\'{e}xico\\
01080, \'{A}lvaro Obreg\'{o}n, M\'{e}xico}
\bigskip

{\it Francisco Estrada \\
Centro de Ciencias de la Atm\'{o}sfera\\
Universidad Nacional Aut\'{o}noma de M\'{e}xico\\
Ciudad Universitaria, 04510, Coyoac\'{a}n, M\'{e}xico}

\bigskip\bigskip
\end{raggedright}  

\section*{Abstract}
Until the beginning of the 1980s, the relation between the Sun and climate 
change was still viewed with suspicion by the wider climate community 
and often remained a ``taboo" subject in the solar astrophysics community. 
The main reason for this fact was a lack of knowledge about the causal link 
between the solar activity and its irradiance, that is, the amount of energy 
received at the average distance between the Earth and the Sun. For many 
years, some authors doubted about the invariability of the solar radiative 
output due to the apparent, but poorly explained, correlations between 
fluctuations of solar activity and atmospheric phenomena.
Research on the mechanisms of solar effects on climate and their 
magnitude is currently benefiting from tremendous renewal of interest. A 
large amount of high resolution data is now available; however, the matter 
remains controversial because most of these records are influenced by other 
factors in addition to solar activity.
In many works, the association between solar and terrestrial environmental 
parameters is found through some type of correlation based on multivariate 
statistics or applying the wavelet analysis on the corresponding time series 
data.
The talk is divided in three parts. In the first I will review the solar-
terrestrial climate problem. Later, I will focus on the time series analysis 
used in our study and finally, I will summarize our preliminary findings and 
comment further ideas to improve our model.

\section*{Review of the solar terrestrial problem}

It is well-known that the Sun has an effect on terrestrial climate since its 
electromagnetic radiation is the main energy for the outer envelopes of 
Earth. This is one of the so called solar-terrestrial physics problems. This 
problem has been the subject of speculation and research by scientists for 
many years. Understanding the behavior of natural fluctuations in the 
climate is especially important because of the possibility of man-induced 
climate changes \citep{DL1971,Schneider1979,Tett1999}.
Many studies have been conducted to show correlations between the solar 
activity and various meteorological parameters but historical observations 
of solar activity were restricted to sunspot numbers and it was not clear 
how these could be physically related to meteorological factors. 
In this section we briefly describe some of the most remarkable 
characteristics of the solar activity and the terrestrial climate, emphasizing 
their possible connection.

\begin{description}
\item[a) Solar Activity]
The term solar activity comprises photospheric and chromospheric 
phenomena such as sunspots, prominences and coronal disturbances. It has 
been measured via satellites during recent decades (for the Total Solar 
Irradiance, TSI) and through other ``proxy" variables in prior times (for 
example, the daily observed number of sunspots, and the concentration of 
some cosmogenic isotopes in ice cores as $^{14}$C and $^{10}$Be).

The phenomena mentioned above are related to the variations of the solar 
magnetic field and the amount of received energy. Those are cyclic 
variations like the 11-year (sunspots),  and 22-year (magnetic field 
reversion), among others.

\item[b) Terrestrial climate]
The Intergovernmental Panel on Climate Change (IPCC) glossary defines 
the climate as the ``average weather," or more rigorously, as the statistical 
description in terms of the mean and variability of relevant quantities (for 
example, temperature, precipitation, and wind)  over a period of time 
ranging from months to thousands or millions of years. 

\item[c) Is there a connection between solar activity \& terrestrial climate?]
There are several hypotheses for how solar variations may affect Earth. 
Some variations, such as changes in Earth's orbit (Milankovitch cycles) are 
only of interest in astronomy. The correlation between cosmic ray fluxes 
and clouds \citep{Laken2010}, as well as, the correlation between the 
number of sunspots and changes in the wind patterns \citep{Willett1949} have 
also been reported. 
Studies about the solar and climate relationship have not been conclusive 
since the actual changes are not enough to account for the majority of the 
warming observed in the atmosphere over the last half of the 20th century.
\end{description}

\noindent
\begin{table}
\caption{Description of the time series for the variables associated to the solar activity and the terrestrial climate.}
\begin{scriptsize}
\begin{tabular}{|m{0.65in}|m{1.1in}|m{0.55in}|m{0.65in}|m{1in}|m{0.5in}|}
\hline
{\bf Description} & {\bf Variable} &{\bf Timescale} &{\bf Stationarity} &{\bf Available period} &{\bf Period of Study} \\
\hline
\multirow{3}{0.65in}{Solar Activity}&Number of sunspots (SP)&daily monthly yearly & No & 1700--2008& \multirow{3}{1in}{1700--1985}\\
\cline{2-5}
                      & Total Solar Irradiation (TSI)  &yearly & No & 1750--1978 (reconstructed time series by \cite{Lean1995} 1978--2008 (satellites)& \\
\cline{2-5}
                      & 10BE concentration in ice cores & geological & No & 1424--1985 & \\
\hline
\multirow{4}{0.65in}{Terrestrial climate}&Global temperature in both hemispheres (TN, TS)&monthly& Breakpoint (1977) & Jan 1850--May 2009&\multirow{4}{0.8in}{Jan 1950 -- May2008} \\ \cline{2-5}
  & Mulivariate ENSO Index (MEI) & monthly & Breakpoint (1977) & Jan 1950--June 2009 & \\ \cline{2-5}
  & North Atlantic Oscillation(NAO) & monthly & Yes & Jul 1821--May 2008 & \\ \cline{2-5}
  & Pacific Decadal Oscillation(PDO) & monthly & Breakpoints (1977,1990) & Jan 1900--Jan 2009 & \\
\hline
\end{tabular}
\end{scriptsize}
\end{table}

\section*{Multivariate Time Series Analysis: VAR methodology}

A common assumption in many time series techniques (e.g. VAR 
methodology) is that the data are stationary. A stationary process has the 
property that the mean, variance and autocorrelation structure do not 
change over time (in other words, a flat looking series without trend, 
constant variance over time, a constant autocorrelation structure over time 
and no periodic fluctuations). If the time series, $Y_{t}$, is not stationary, we can 
often transform it with one of the following techniques: 1) apply a Box-Cox 
transformation (logarithm is the simplest), and/or 2) differentiate the data 
$Y_{t}$ to create a new series $X_{t}$
($X_{t}$ = $\nabla Y_{t}$ = $Y_{t}$--$Y_{t-1}$ ;
$X_{t}$ =$\nabla^{2} Y_{t}$ = $Y_{t}$ -- 2$Y_{t-1}$ + $Y_{t-2}$).

\subsection*{VAR methodology: description}
The Vector AutoRegression (VAR) model is one of the most successful,
flexible, and easy to use models for the analysis of multivariate time series.
It is a natural extension of the univariate autoregressive model to dynamic
multivariate time series \citep{Sims1972,Sims1980,Sims1982}. It describes the evolution
of a set of $k$ variables (called endogenous variables) over the same sample
period (t = 1, ..., T) as a linear function of only their past evolution:
\begin{equation} 
y_{t}=c+\alpha_{1}y_{t-1}+\alpha_{2}y_{t-2}+...+\alpha_{p}y_{t-p}+\epsilon_{t}
\end{equation} 
where $c$ is a $k \times 1$ vector of constants (intercept) , $\alpha_{i}$
is a $k \times k$ matrix (for every i=1,.., p), p is the number of lags
(that is, the number of periods back), and $\epsilon_{t}$ is a $k \times 1$
vector of error terms. There are some additional
assumptions about the error terms: 1) the expected value is zero, that is,
E[$\epsilon_{it}$]=0 with t=1,..., T, and 2) the errors are not autocorrelated,
E[$\epsilon_{it}$ $\epsilon_{jt}$]=0   with $t \neq \tau$.

The determination of the number of lags p is a trade-off between the
dimensionality and abbreviate models. To find the optimal lag length we
can apply a Log-Likelihood Ratio test (LR) test or an information criterion
\citep{L1993}.

Once the estimation of the parameters in the VAR(p) model shown in Eq.
(1) through Ordinary Least Squares (OLS), we need to interpret the
dynamic relationship between the indicated variables using the Granger
causality.

\subsection*{Application of the VAR methodology to model the solar activity and
terrestrial climate connection}
Our purpose is to investigate the relationship between the solar activity and
the major climate phenomena by means of time series analysis. The data
are taken from the National Geophysical Data Center (NGDC) and the
National Climatic Data Center (NCDC). In Table 1 we summarize the
selected variables for each physical system.

\begin{description}
\item[a) VAR model for the solar-terrestrial climate connection]
We estimate a VAR(p) model using the variables shown in Table 1
where the non-stationary time series have been differentiated once. The
exogenous variables are: a) number of sunspots, b) TSI, c) d76 (dummy
variable for 1976), d) d77 (dummy variable for 1977) and e) d90
(dummy variable for 1990).

The optimal lag length is 4 and the VAR(4) is formed by 5 equations
(TN, TS, MEI, NAO, and PDO). The statistical validation is shown in
Table 2.

\item[b) VAR model for the solar activity]
We estimate a VAR(p) model using the variables shown in Table 1
where the non-stationary time series have been differentiated once.

The optimal lag length is 8 and the VAR(8) is formed by 3 equations
(SP, TSI, and $^{10}$BE). The statistical validation is shown in Table 2.
\end{description}

\noindent
\begin{table}
\begin{small}
\caption{Description of the time series for the variables associated to the solar activity and the terrestrial climate.}
\begin{tabular}{|m{1.6in}|m{1.7in}|m{1.5in}|}
\hline
\multirow{2}{1.6in}{\bf VAR(p) characteristic} & \multicolumn{2}{c|}{Model} \\ \cline{2-3}
 & Solar-terrestrial \ \  climate connection & Solar activity \\ \hline
Lag length (p) &  4 & 8\\ 
\hline
Significance of the coefficients & All (except equation for NAO) & All \\
\hline
Stability & Yes & Yes \\
\hline
Homoskedasticity of the residuals & No & No \\
\hline
Normality of the residuals & Yes (except equation for TN) & No \\
\hline
Granger causality & TSI does not affect MEI, NAO and PDO & 10BE does not affect SP and TSI \\
\hline
\end{tabular}
\end{small}
\end{table}

\section*{Summary and ideas for future work}

The possible relation between the solar activity and the terrestrial climate
has been addressed in many works. Most of them search for periodicities or
correlations among the set of variables that characterize the solar activity
and the main climate parameters. For example, the ``wavelet analysis"
cannot be the most adequate to analyze multivariate time series.
In this work we have proposed and estimated a VAR model to explain such
a connection. For this model we have analyzed the time series for the most
remarkable characteristics of both the solar activity and climate. Our main
results and some ideas for future work are listed below.

\begin{itemize}

\item The solar activity is modeled by a VAR(8) in which we find that the $^{10}$Be
concentration does not play a fundamental role.

\item The solar activity and terrestrial climate connection is modeled by a VAR(4) where
the solar variables are taken as exogenous. It seems that the sun (described only for
the number of sunspots and the TSI) has a weak connection to Earth, at least for the
major climate phenomena.

\item It is convenient to include a term related to the cloudiness to verify the previous
findings.
\item Analyzing certain cycles in the solar activity could help us to determine the epochs
in which the connection with the terrestrial climate was stronger.

\item We need to search for other proxy variables that describe the solar activity and
introduce variables related to regional climate (for example: precipitation, pressure,
local temperature).

\end{itemize}

\section*{Acknowledgements}
E. Mart\'{i}nez-G\'{o}mez thanks to the Faculty For The Future Program and
CONACyT-Mexico postdoctoral fellowship for their financial support to this
research. V. M. Guerrero acknowledges support from Asociaci\'{o}n Mexicana
de Cultura, A. C.

\newpage
\setcounter{section}{0}
\setcounter{figure}{0}

\Title{Optimal Scheduling of Exoplanet Observations Using Bayesian Adaptive Exploration}
\bigskip\bigskip
\begin{raggedright}  
{\it Thomas J. Loredo\\
Department of Astronomy\\
Cornell University\\
Cornell, New York, USA}
\bigskip\bigskip
\end{raggedright}

This presentation describes ongoing work by a collaboration of
astronomers and statisticians developing a suite of Bayesian tools for
analysis and adaptive scheduling of exoplanet host star reflex motion
observations.  In this presentation I  focus on the most unique aspect
of our work:  adaptive scheduling of observations using the principles of
Bayesian experimental design in a sequential data analysis setting.  I 
introduce the core ideas and highlight some of the computational challenges
that arise when implementing Bayesian design with nonlinear models. 
Specializing to parameter estimation cases (e.g., measuring the orbit of
planet known to be present), there is an important simplification that
enables relatively straightforward calculation of greedy designs via maximum
entropy (MaxEnt) sampling.  We implement MaxEnt sampling using
population-based MCMC to provide samples used in a nested Monte Carlo
integration algorithm.  I demonstrate the approach with a toy problem, and
with a re-analysis of existing exoplanet data supplemented by simulated
optimal data points.

Bayesian adaptive exploration (BAE) proceeds by iterating a three-stage cycle:
{\em Observation--Inference--Design}.  Figure~1 depicts the flow of
information through one such cycle.  In the observation stage, new data are
obtained based on an observing strategy produced by the previous cycle of
exploration.  The inference stage synthesizes the information provided by
previous and new observations to produce interim results such as signal
detections, parameter estimates, or object classifications.  In the
design stage the interim inferences are used to predict future data for a
variety of possible observing strategies; the strategy that offers the
greatest expected improvement in inferences, quantified with
information-theoretic measures, is passed on to the next
Observation--Inference--Design cycle.

\begin{figure}
\centerline{\includegraphics[width=\textwidth]{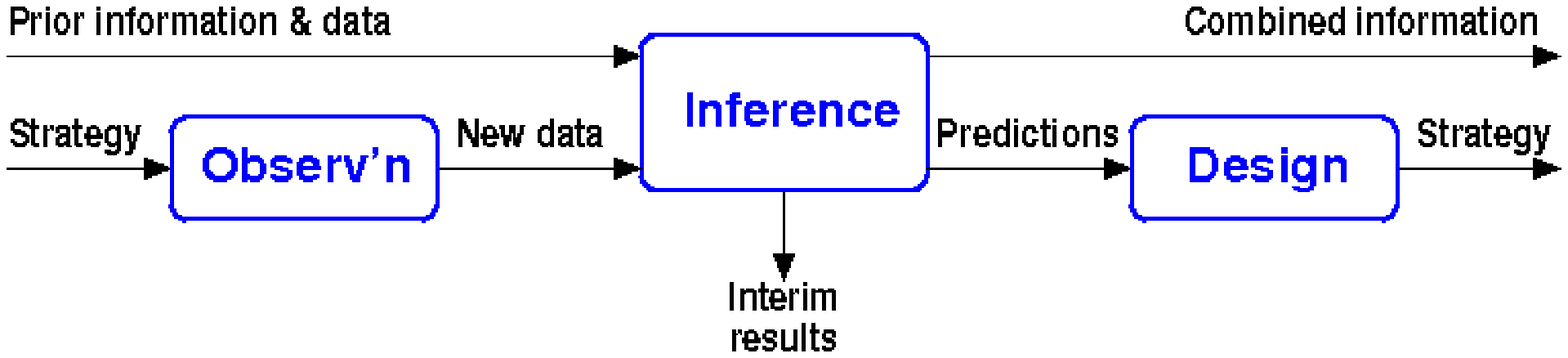}}
\caption{Depiction of one cycle of the three-stage Bayesian adaptive
exploration process.}
\end{figure}

Figures 2 and 3 show highlights of application of BAE to radial velocity
(RV) observations of the single-planet system HD~222582.  \cite{Vogt2000}
reported 24 observations obtained over a 683~d time span with
instrumentation at the Keck observatory; \cite{Butler2006} reported 13
subsequent observations.  We consider the early observations as a starting
point, and compare inferences based on simulated subsequent data at optimal
times identified via BAE to inferences using the actual, non-optimal
subsequent data (we generated simulated data using the best-fit orbit for
all 37 actual observations). Figure~2 shows how the optimal observing time
for the first new datum is calculated.  Bayesian analysis based on the 24
early data points (red diamonds) produces a posterior distribution for the
orbital parameters.  We explore the posterior via population-based adaptive
Markov chain Monte Carlo sampling, producing an ensemble of possible orbits.

\begin{figure}
\centerline{\includegraphics[width=\textwidth]{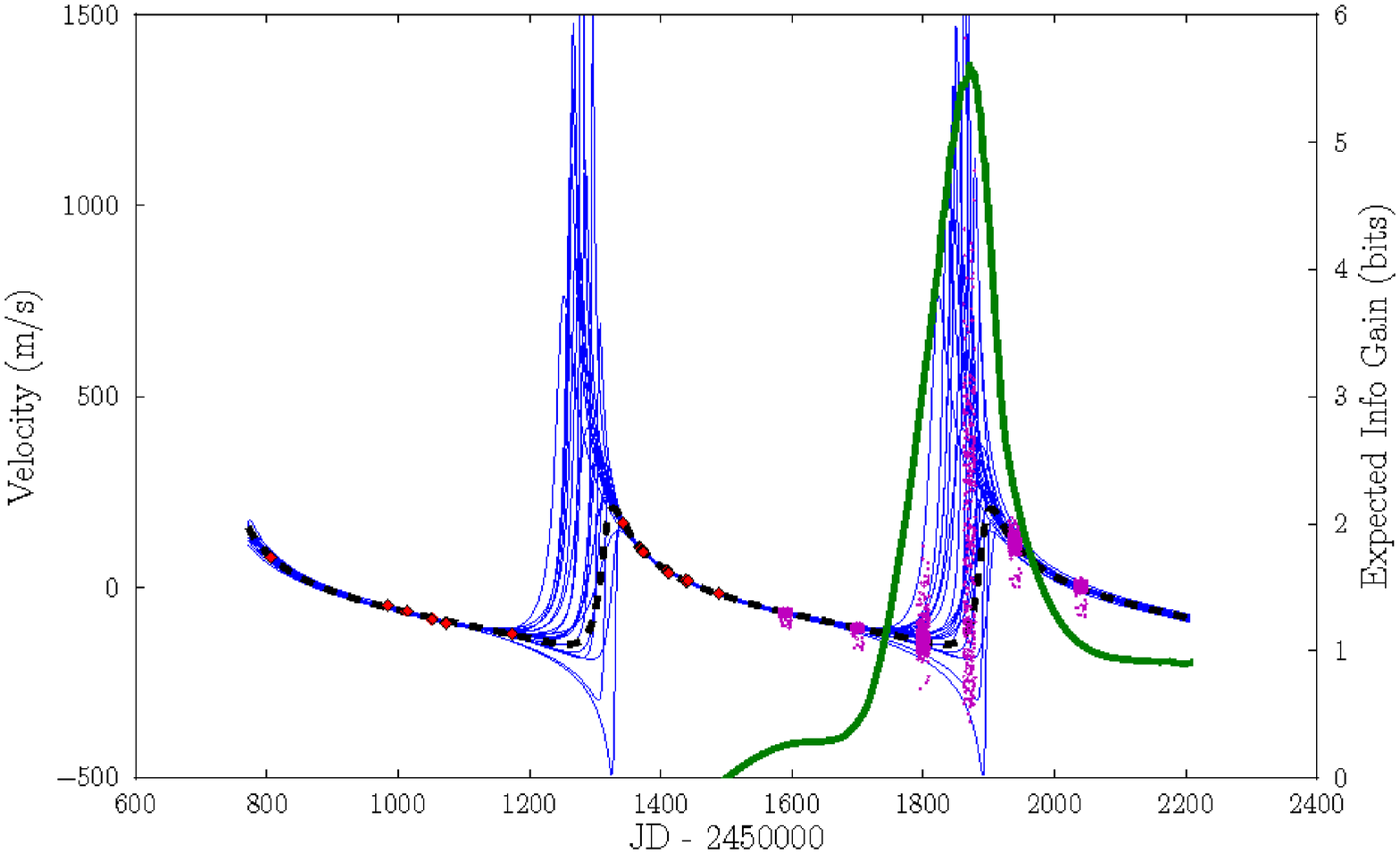}}
\caption{Results based on data from HD~222582:  Observed (red diamonds) and
predicted (ensemble of thin blue curves; also magenta dots at selected
times) velocity vs.\ time (against left axis); entropy of predictive
distribution vs.\ future observing time (green curve, against right axis).}
\end{figure}

 The blue curves show the velocity curves for 20 posterior samples, roughly
depicting the predictive distribution for future data vs.\ time; the magenta
point clouds provide a more complete depiction at selected times, showing
$\sim 100$ samples from the predictive distribution at six future times. For
this type of data, the expected information gain from future data is
proportional to the entropy (uncertainty) in the predictive distribution,
so one expects to learn the most by observing where the predictions are most
uncertain.  The green curve (against the right axis) quantifies this,
showing the entropy in the predictive distribution vs.\ time (in bits
relative to repeating the last actual observation), calculated using the
predictive distribution samples; its peak identifies the optimal observing
time.  We generated a single new observation at this time, and repeated the
procedure twice more to produce three new optimal observations. The left
panel of Figure~3 shows inferences (samples and histograms for marginal
distributions) based on the resulting 27 observations; the right panel shows
inferences using the 37 actual observations.  Inferences with the fewer but
optimized new observations are much more precise.

\begin{figure}
\includegraphics[width=0.5\textwidth]{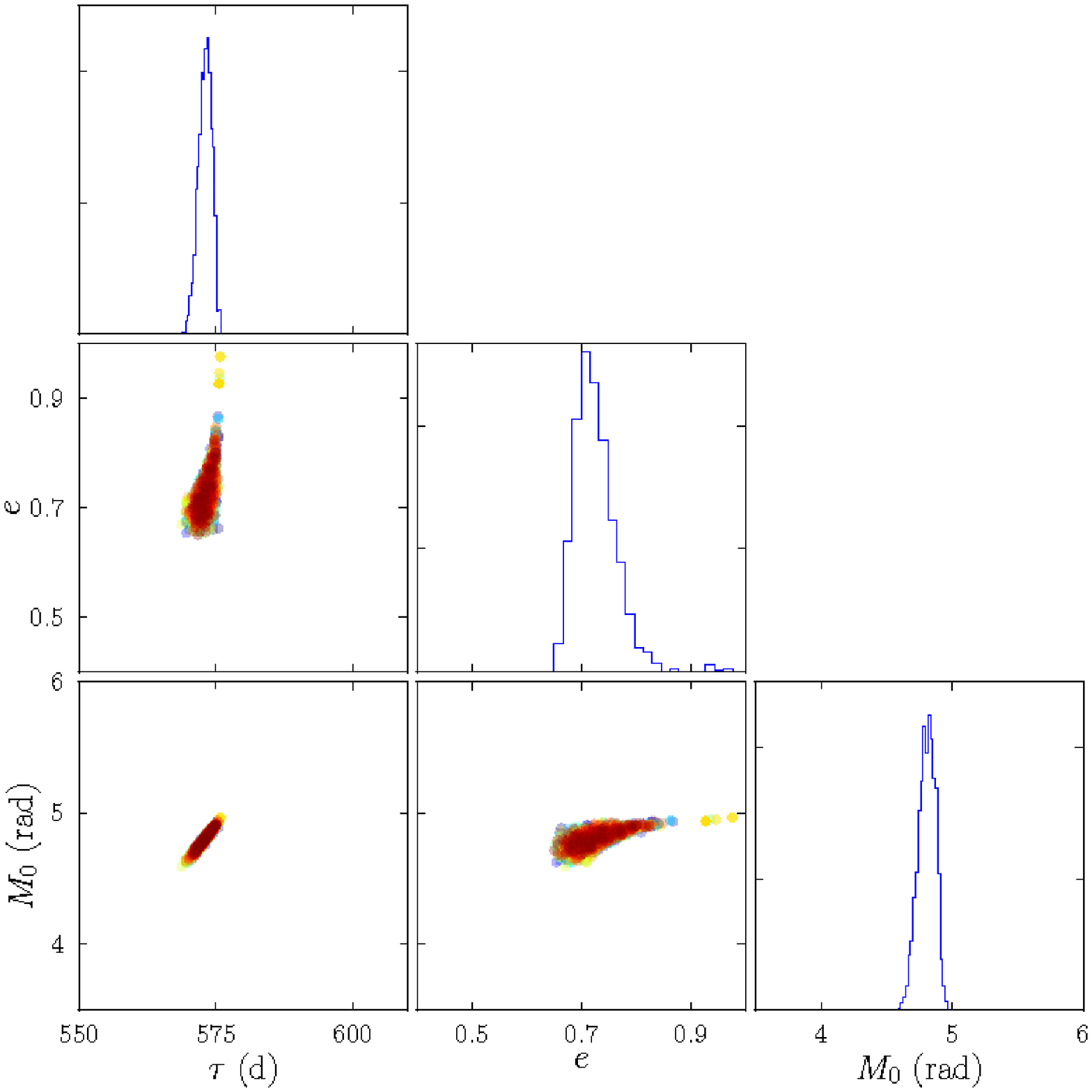}
\includegraphics[width=0.5\textwidth]{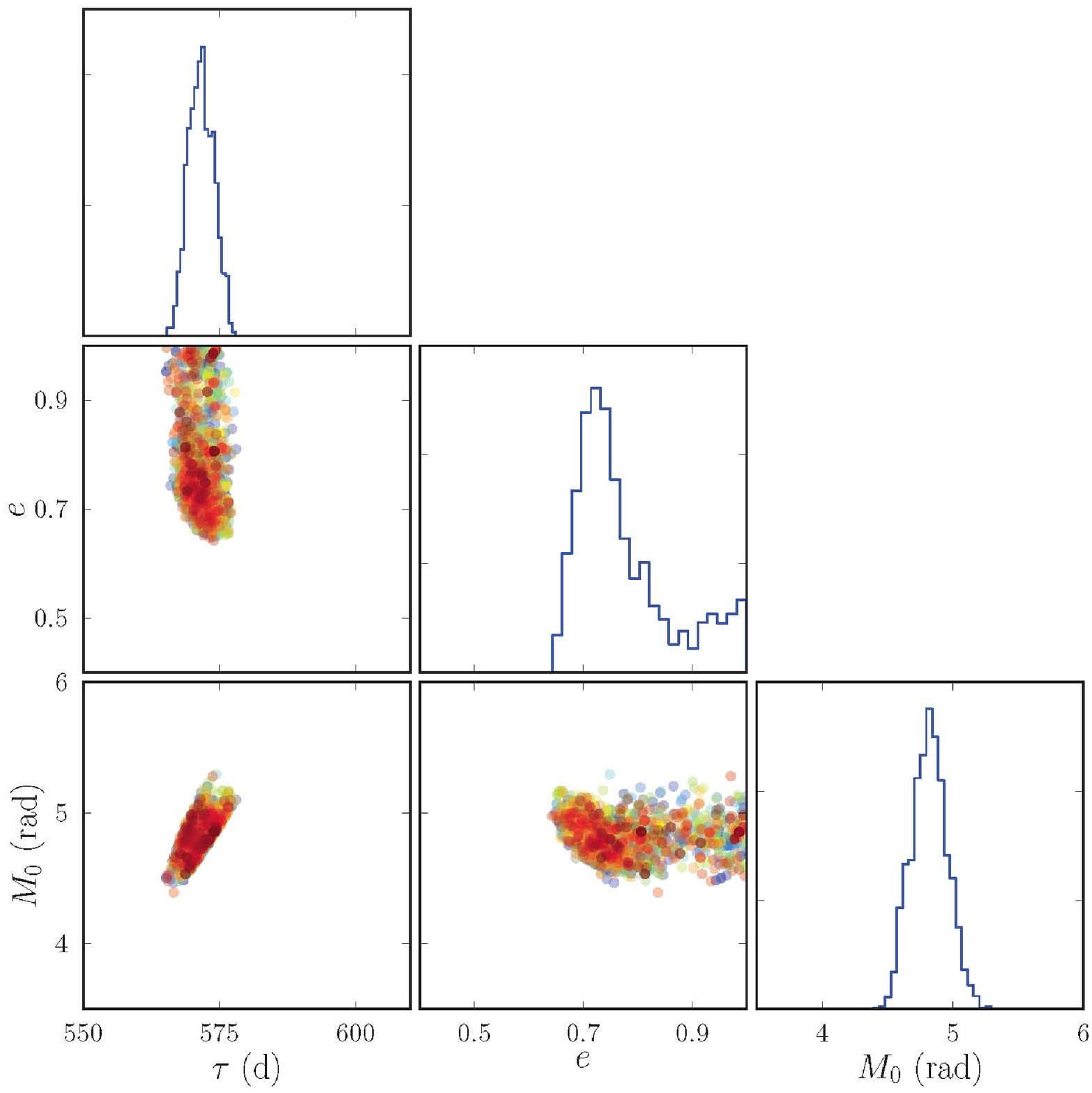}
\caption{Orbital parameter estimates based on 24 early
observations and three simulated new observations at optimal times (left),
and based on 24 early and 13 new, non-optimal actual observations (right).
Parameters are period, $\tau$, eccentricity, $e$, and mean anomaly
at a fiducial time, $M_0$.}
\end{figure}

\newpage
\setcounter{section}{0}
\setcounter{figure}{0}

\Title{Beyond Photometric Redshifts using Bayesian Inference}
\bigskip\bigskip
\begin{raggedright}  
{\it  Tam\'{a}s Budav\'{a}ri\\
Johns Hopkins University\\
Baltimore, Maryland, USA}
\bigskip\bigskip
\end{raggedright}

The galaxies in our expanding Universe are seen redder than their actual
emission. This redshift in their spectra are typically measured from narrow
spectral lines by identifying them to known atomic lines seen in the lab.
Spectroscopy provides accurate determinations of the redshift as well as a great
insight into the physical processes but is very expensive and time consuming.
Alternative methods have been explored for decades.
The field of photometric redshifts started when Baum (1962) first compared the
magnitudes of red galaxies in distant clusters to local measurements. The first
studies were colorful proofs of the concept, which demonstrated the adequate
precision of estimating galaxy redshifts based on photometry alone, without
spectroscopic follow up. The new field became increasingly important over time,
and with the upcoming multicolor surveys just around the corner, the topic is
hotter than ever.
Traditional methods can be broken down into two distinct classes: empirical and
template fitting. 
Empirical methods rely on training sets of objects with known photometry and
spectroscopic redshifts. The usual assumption is that galaxies with the same
colors are at identical redshifts. The redshift of a new object is derived based
on its vicinity in magnitude space to the calibrators of the training set.
Polynomial fitting, locally linear regression and a plethora of other machine
learning algorithms have been tested and, usually, with good results. They are
easy to implement, fast to run, but are only applicable to new objects with the
same photometric measurements in the same regime.
Template fitting relies on high-resolution spectral models, which are
parameterized by their type, brightness and redshift. The best fitting parameters
are typically sought in a maximum likelihood estimation. It is simple to
implement and work for new detections in any photometric system but the results
are only as good as the template spectra.

\begin{figure}[!htb]
\includegraphics[scale=0.65]{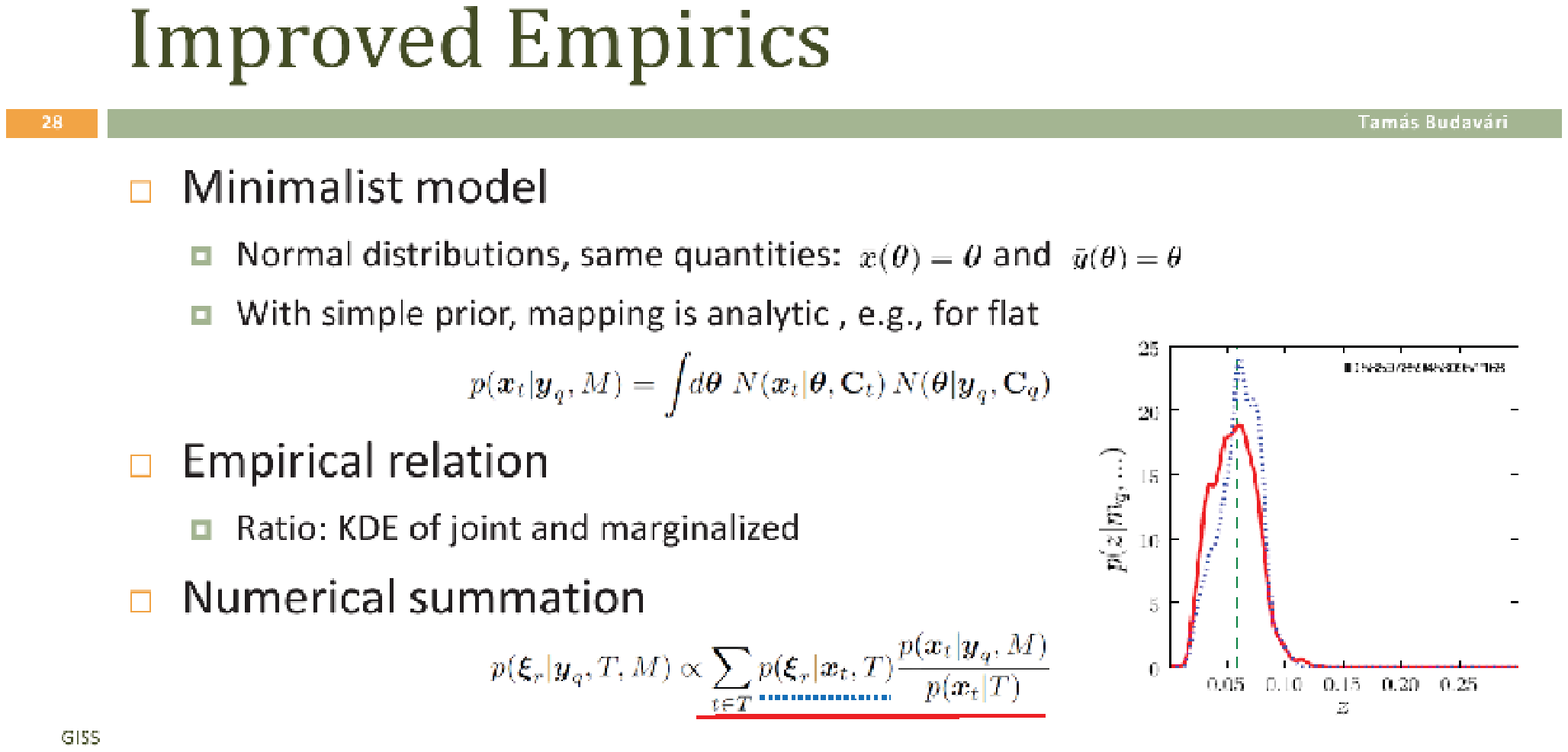}
\caption{The presented slide illustrates the minimalist model to incorporate
the photometric uncertainties of the new galaxy. Using KDE the relation is 
estimated (blue dotted line), which is averaged with the uncertainty for 
the final result (red solid line).}
\label{budavari-fig01}
\end{figure}

To understand the implications of the previous methods and to point toward more
advanced approaches, we can use Bayesian inference (Budav\'ari 2009). Constraints
on photometric redshifts and other physical parameters in the more general
inversion problem are derived from first principles. We combine two key
ingredients:\par
(1) Relation to physics --- The redshift (z) is not simply a function of the
observables. Considering that they cannot possibly capture all the information,
one expects a spread. Regression is only a good model, when the width is very
narrow. Otherwise one should estimate the conditional density of $p(z|m)$. While
not easy in practice, this is conceptually straightforward to do on a given set
of calibrators.\par
(2) Mapping of the observables --- If we would like to perform the estimation of
a new object with colors in a separate photometric system, its $m'$ magnitudes
need to be mapped on to the passbands of the training set (m). This might be as
simple as an empirical conversion formula or as sophisticated as  spectral
synthesis. In general, a model is needed, which can also propagate the
uncertainties, $p(m|m')$. The final density is the convolution of these two
functions.\par

After 50 years of pragmatism, photometric redshift estimation is now placed on a
firm statistical foundation. We can put the traditional methods in context; they
are special cases of a unified framework. Their conceptual weaknesses become
visible from this aspect. The new approach points us toward more advanced methods
that combine the advantages of the earlier techniques. In addition we can
formally learn about selecting calibration sets for specific studies. These
advancements are going to be vital for analyzing the observations of the
next-generation survey telescopes.

\newpage
\setcounter{section}{0}
\setcounter{figure}{0}

\Title{Long-Range Climate Forecasts Using Data Clustering and Information
Theory}
\bigskip\bigskip
\begin{raggedright}  
{\it Dimitris Giannakis\\
New York University\\
New York, New York, USA}
\bigskip\bigskip
\end{raggedright}

Even though forecasting the weather beyond about two weeks is not possible,
certain climate processes (involving, e.g., the large-scale circulation in the
Earth's oceans) are predictable up to a decade in advance. These so-called
climate regimes can influence regions as large as the West Coast of North America
over several years, and therefore developing models to predict them is a problem
of wide practical impact. An additional central issue is to quantify objectively
the errors and biases that are invariably associated with these models. 

In classical studies on decadal prediction \citep[][]{Boer00,Boer04,Collins02},
ensemble experiments are performed using one or more climate models initialized
by perturbed initial conditions relative to a reference state, and predictive
skill is measured by comparing the root mean square difference of ensemble
trajectories to its equilibrium value. However, skill metrics of this type have
the drawback of not being invariant under invertible transformations of the
prediction variables, and not taking into account the important issue of model
error. Further challenges concern the choice of initial conditions of the
ensemble members \citep[][]{HurrellEtAl09,MeehlEtAl09,SolomonEtAl09}.

In this talk, we present recent work \citep[][]{GiannakisMajda11a,GiannakisMajda11b}
on methods based on data clustering and information theory to build and assess
probabilistic models for long-range climate forecasts that address several of
the above issues. The fundamental perspective adopted here is that predictions
in climate models correspond to transfer of information; specifically transfer
of information between the initial conditions (which in general are not known
completely) and the state of the climate system at some future time. This opens
up the possibility of using the mathematical framework of information theory to
characterize both dynamical prediction skill and model error with metrics that
are invariant under invertible nonlinear transformations of observables
\citep[][]{Kleeman02,RoulstonSmith02,MajdaEtAl02b,MajdaEtAl05,DelSoleTippett07,MajdaGershgorin10}. 

The key points of our discussion are that (i) the long-range predictive skill of
climate models can be revealed through a suitable coarse-grained partition of the
set of initial data available to a model (which are generally incomplete);
(ii) long-range predictive skill with imperfect models depends simultaneously
on the fidelity of these models at asymptotic times, their fidelity during
dynamical relaxation to equilibrium, and the discrepancy from equilibrium of
forecast probabilities at finite lead times. Here, the coarse-grained partition
of the initial data is constructed by data-clustering equilibrium realizations
of ergodic dynamical systems without having to carry out ensemble
initializations. Moreover, prediction probabilities conditioned on the clusters
can be evaluated empirically without having to invoke additional assumptions
(e.g., Gaussianity), since detailed initial conditions are not needed to sample
these distributions. 

In this framework, predictive skill corresponds to the additional information
content beyond equilibrium of the cluster-conditional distributions. The natural
information-theoretic functional to measure this additional information is
relative entropy, which induces a notion of distance between the
cluster-conditional and equilibrium distributions. A related analysis leads to
measures of model error that correspond to the lack of information (or ignorance)
of an imperfect model relative to the true model. The techniques developed here
have potential applications across several disciplines involving dynamical system
predictions.

As a concrete application of our approach, we study long-range predictability in
the equivalent barotropic, double-gyre model of \citet[][]{McCalpinHaidvogel96}
(frequently called the ``1.5-layer model''). This simple model of ocean
circulation has non-trivial low-frequency dynamics, characterized by infrequent
transitions between meandering, moderate-energy, and extensional configurations
of the eastward jet (analogous to the Gulf Stream in the North Atlantic). The
algorithm employed here for phase-space partitioning involves building a
multi-time family of clusters, computed for different temporal intervals of
coarse graining; a recipe similar to kernel density estimation methods. We
demonstrate that knowledge of cluster affiliation in the computed partitions
carries significant information beyond equilibrium about the total energy and
the leading two principal components (PCs) of the streamfunction (which are
natural variables for the low-frequency dynamics of this system) for five- to
seven-year forecast lead times, i.e., for a timescale about a factor of five
longer than the maximum decorrelation time of the PCs.

As an application involving imperfect models, we discuss the error in Markov
models of the switching process between the ocean circulation regimes. Imposing
Markovianity on the transition process is a familiar approximation in this
context \citep[][]{FranzkeEtAl08,FranzkeEtAl09}, though the validity of this
assumption typically remains moot. Our analysis exposes starkly the falseness of
predictive skill that one might attribute to a Markovian description of the
regime transitions in the 1.5-layer model model by relying on an (internal)
assessment based solely on the deviation of the time-dependent prediction
probabilities of the Markov model from its biased equilibrium. In particular,
we find that a Markov model associated with a seven-state partition appears to
outperform a three-state model, both in its discriminating power and its
persistence (measured respectively by the deviation from equilibrium and rate
of approach to equilibrium), when actually the skill of the seven-state model
is false because its equilibrium statistics are biased. Here, the main conclusion
is that evaluating simultaneously model errors in both the climatology and the
dynamical relaxation to equilibrium should be an integral part of assessments of
long-range forecasting skill.

\vspace{1ex}

\noindent\emph{Acknowledgment} The material presented in this talk is joint
work with Andrew Majda of New York University. This research is partially
supported by NSF grant DMS-0456713, from ONR DRI grants N25-74200-F6607 and
N00014-10-1-0554, and from DARPA grants N00014-07-10750 and N00014-08-1-1080.


\newpage
\setcounter{section}{0}
\setcounter{figure}{0}

\Title{Comparison of Information-Theoretic Methods to estimate the information
flow in a dynamical system} 
\bigskip\bigskip

\begin{raggedright}  
{\it Deniz Gencaga\\
NOAA Crest\\
New York, New York, USA}
\bigskip\bigskip
\end{raggedright}

\section*{Abstract}
In order to quantify the amount of information about a variable, or to quantify
the information shared between two variables, we utilize information-theoretic
quantities like entropy and mutual information (MI), respectively. If these
variables constitute a coupled, dynamical system, they share the information
along a direction, i.e. we have an information flow in time. In this case, the
direction of the information flow also needs to be estimated. However, MI does
not provide directionality. Transfer entropy (TE) has been proposed in the
literature to estimate the direction of information flow in addition to its
magnitude. Here, our goal is to estimate the transfer entropy from observed
data accurately. For this purpose, we compare most frequently used methods in
the literature and propose our own technique. Unfortunately, every method has
its own free tuning parameter(s) so that there is not a consensus on an optimal
way of estimating TE from a dataset. In this work, we compare several methods
along with a method that we propose, on a Lorenz model. Here, our goal is to
develop the required appropriate and reliable mathematical tool synthesizing
from all of these disjoint methods used in fields ranging from biomedicine to
telecommunications and apply the resulting technique on tropical data sets to
better understand events such as the Madden Julian Oscillation in the future.

\section*{Introduction and problem statement}

Nonlinear coupling between complex dynamical systems is very common in many
fields \citep{wallace2006,gourvitch2007}. In order to study the relationships
between such coupled
subsystems, we must utilize higher-order statistics. Thus, using linear
techniques based on correlation analysis cannot be sufficient. Although using
mutual information seems to be a good alternative to model higher order
nonlinear relationships, it becomes insufficient when we would like to model
the directionality of the interactions between the variables, as it is a
symmetric measure. Thus, \cite{schrieber2000}, proposed an asymmetric measure
between two variables X and Y as follows to study directional interactions:

\begin{equation}
    \label{simple_equation1}
    {TE}_{{Y}\rightarrow{X}}={T}\left({X}_{i+1}\mid{{\bf X_i^{(k)}},{\bf Y_j^{(l)}}}\right)= \sum_{i=1}^Np\left(x_{x_{i+1}},{\bf x_i^{(k)}},{\bf y_j^{(l)}} \right)log_2\frac{  p\left(x_{i+1}\mid{ {\bf x_i^{(k)}           },{\bf y_j^{(l)}              } }\right)       }{p\left(x_{i+1}    \mid{       { \bf  x_i^{(k)}                             }      }   \right)}
    \end{equation}
    
    \begin{equation}
    \label{simple_equation2}
    {TE}_{{X}\rightarrow{Y}}={T}\left({Y}_{i+1}\mid{{\bf Y_i^{(k)}},{\bf X_j^{(l)}}}\right)= \sum_{i=1}^Np\left(y_{y_{i+1}},{\bf y_i^{(k)}},{\bf x_j^{(l)}} \right)log_2\frac{  p\left(y_{i+1}\mid{ {\bf y_i^{(k)}           },{\bf x_j^{(l)}              } }\right)       }{p\left(y_{i+1}    \mid{       { \bf  y_i^{(k)}                             }      }   \right)}
    \end{equation}

where $ x_i(k)=[x_i,\ldots,x_{i-k+1}]$ and $y_j(l)=[y_i,\ldots,y_{i-l+1}]$ are
past states and X and Y are $k^{th}$ and $l^{{th}}$ order Markov processes,
respectively.
Above, $ {TE}_{{X}\rightarrow{Y}}$  and  ${TE}_{{Y}\rightarrow{X}}$ denote
transfer entropies in the direction from X to Y and from Y to X, respectively.
For example, in Equation \ref{simple_equation1}, TE describes the degree
to which, information
about Y allows one to predict future values of X. This is a causal influence
that the subsystem Y has on the subsystem X. 

\subsection*{Estimation}

In the literature, there are a couple of methods to estimate this quantity from
data. However, they have their own fine tuning parameters so that there is not
a consensus on an optimal way of estimating TE from a dataset. One of them is
the Kernel Density Estimation method \citep{sabeson2007}, where an
optimal radius needs to be picked. In another method, called the Adaptive
Partitioning of the observation space \citep{DV1999}, we make use
of the unequal bin sized histograms for
mutual information estimations. However, as we have to subtract multivariate
mutual information quantities to estimate the final TE, this could be affected
by biases. Thus, we propose our own method of generalized Bayesian
piecewise-constant model for the probability density function estimation and
then calculate TE using the summation formula of individual Shannon entropies. 

\begin{figure}
    \includegraphics[width=1\textwidth]{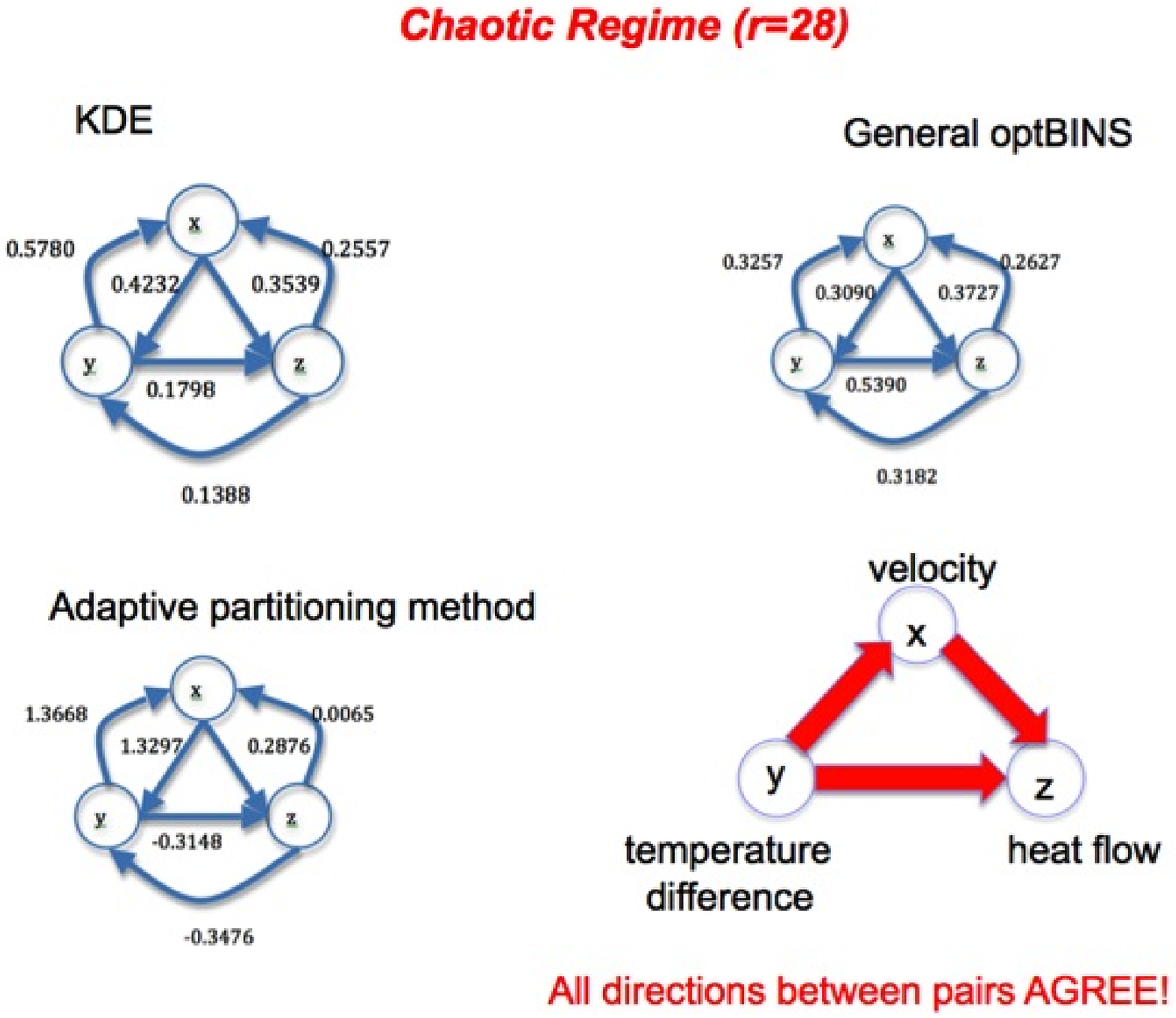}
    \caption{Estimated transfer entropies between the pairs of Lorenz equations using three different methods}
    \label{simulationfigure}
\end{figure}

\section*{EXPERIMENT AND CONCLUSION}

We tested three methods on the nonlinear coupled Lorenz equations given below.
In climatology, they represent a model of an atmospheric convection roll where
x, y, z denote convective velocity, vertical temperature difference and mean
convective heat flow, respectively. Using the three methods, we obtained the
TE estimates illustrated in Figure \ref{simulationfigure}.
\begin{equation}
    \label{simple_equation3}
     \frac{dx}{dt}= \sigma \left( {y-x} \right)
\end{equation}
\begin{displaymath}
    \label{simple_equation4}
     \frac{dy}{dt}={-xz+rx-y}
\end{displaymath}
\begin{displaymath}
    \label{simple_equation5}
     \frac{dz}{dt}={xy-bz}
\end{displaymath}

where $\sigma=10$, $b=\frac{8}{3}$, r: Rayleigh number and $r=28$ in a chaotic regime.

In conclusion, computer simulations demonstrate that we can find a reliable
parameter regime for all methods at the same time and estimate TE direction
from data so that we can identify the information flow between the variables
reliably, as all methods agree both mathematically and physically. Currently,
we are working on the magnitude differences between the methods so that we can
apply TE to identify the information flow between the relevant climate
variables of the tropical disturbance, called Madden Julian Oscillation (MJO).
Later, this technique will be developed for us to better understand the highly
nonlinear nature of atmospheric dynamics.

\newpage
\setcounter{section}{0}
\setcounter{figure}{0}

\Title{Reconstructing the Galactic halo's accretion history: A finite
mixture model approach}
\bigskip\bigskip
\begin{raggedright}  

{\it Duane Lee\\
Department of Astronomy\\
Columbia University\\
New York, New York, USA}

\bigskip

{\it Will Jessop\\
Department of Statistics\\
Columbia University\\
New York, New York, USA}

\bigskip\bigskip
\end{raggedright}  

\section*{Abstract}

The stellar halo that surrounds our Milky Way Galaxy is thought to
have been built, at least in part, from the agglomeration of stars
from many smaller galaxies. This talk outlines an approach to
reconstructing the history of Galactic accretion events by looking at
the distribution of the chemical abundances of halo stars. The full
distribution is assumed to result from the superpositionof stellar
populations accreted at different times from progenitor galaxies of
different sizes. Our approach uses the Expectation-Maximization (EM)
algorithm to find the maximum-likelihood estimators that assess the
contribution of each of these progenitors in forming the Galaxy.
\newpage
\setcounter{section}{0}
\setcounter{figure}{0}

\begin{center}
{\Large {\bf Scientific Program}}
\end{center}
\bigskip

\begin{footnotesize}
\noindent
\begin{tabular}{lp{2.7in}l}
           & {\bf Thursday, February 24, 2011}& \\
& & \\
{\it 10:00}& {\it Welcome  }& Michael Way (GISS)\\
           &                & Catherine Naud (GISS)\\
10:20     &  How long will it take. A historical approach to boundary crossing.
& Victor de la Pe\~{n}a (Columbia)\\
10:45     &  Cosmology through the Large-Scale Structure of the Universe
& Eyal Kazin (NYU)\\
11:30     &  On the shoulders of Gauss, Bessel, and Poisson: links, chunks,
spheres, and conditional models
& William Heavlin (Google)\\
12:15     &  Mining Citizen Science Data: Machine Learning Challenges
& Kirk Borne (GMU)\\

13:00     & {\it Lunch Break} & \\

14:30      & Tracking Climate Models: Advances in Climate Informatics
& Claire Monteleoni (Columbia)\\
15:15      & Spectral Analysis Methods for Complex Source Mixtures
& Kevin Knuth (SUNY/Albany)\\
16:00      & Beyond Objects: Using machines to understand the diffuse universe
& Joshua Peek (Columbia)\\
16:45      & Viewpoints: A high-performance visualization and analysis tool
& Michael Way (GISS)\\
           &   & \\
           & {\bf Friday, February 25, 2011} & \\
           &   & \\
10:00      & Clustering Approach for Partitioning Directional Data in Earth and Space Sciences
& Christian Klose (Think Geohazards)\\
10:45      & Planetary Detection: The Kepler Mission
& Jon Jenkins (NASA/Ames)\\
11:30      & Understanding the possible influence of the solar activity on the
terrestrial climate: A times series analysis approach
& Elizabeth Mart\'{i}nez-G\'{o}mez (Penn State)\\
12:15      & Bayesian adaptive exploration applied to optimal scheduling of
exoplanet Radial Velocity observations
& Tom Loredo (Cornell)\\
13:00      & {\it Lunch Break} & \\
14:30      & Bayesian Inference from Photometric Surveys
& Tam\'{a}s Budav\'{a}ri (JHU)\\
15:15      & Long-Range Forecasts Using Data Clustering and Information Theory
& Dimitris Giannakis (NYU)\\
16:00      & Comparison of Information-Theoretic Methods to estimate the
information flow in a dynamical system
& Deniz Gencaga (CCNY)\\
16:45      & Reconstructing the Galactic halo's accretion history: A finite
mixture model approach
& Duane Lee and Will Jessop (Columbia)\\
\end{tabular}
\end{footnotesize}

\newpage
\setcounter{section}{0}
\setcounter{figure}{0}

\begin{center} {\Large {\bf Participants}} \end{center}
\begin{small}
\begin{tabular}{|p{1.4in}|p{3.0in}|p{0.5in}|}
\hline
Ryan Abrahams &CUNY Graduate Center& \\
Hannah Aizenman &CCNY/Computer Science& \\
Arif Albayrak &NASA GES DISC& \\
Kehinde-Paul Alabi &CCNY/Computer Science& \\
\hline
Vivienne Baldassare &CUNY/Hunter College/Physics& \\
Imre Bartos &Columbia/Physics& \\
Mike Bauer &Columbia-NASA/GISS& \\
Rohit Bhardwaj &Columbia/Computer Science& \\
Amitai Bin-Nun &University of Pennsylvania/Astronomy& \\
James Booth &NASA/GISS& \\
Kirk Borne &George Mason University& speaker\\
Andreas Breiler &IdeLaboratoriet& \\
Douglas Brenner &American Museum of Natural History& \\
Tamas Budavari &Johns Hopkins University& speaker\\
\hline
Samantha Chan &Lamont-Doherty Earth Observatory, Columbia University& \\
Milena C. Cuellar &BCC. CUNY& \\
Victor de al Pena &Columbia/Statistics& speaker\\
\hline
Danielle Dowling &CUNY Graduate Center& \\
Tarek El-Gaaly &Rutgers/Computer Science& \\
Daniel Feldman &CUNY/College of Staten Island& \\
Stephanie Fiorenza &CUNY/College of Staten Island/Physics& \\
Carl Friedberg &NYC/Comets.com& \\
\hline
Deniz Gencaga &CCNY/EE, NOAA-CREST& speaker\\
Dimitrios Giannakis &Center for Atmosphere Ocean Science, Courant Institute of Mathematical Sciences& speaker\\
Irina Gladkova &CCNY/Computer Science& \\
Elizabeth Gomez &Penn State University/Statistics&speaker\\
Joshua Gordon &Columbia/Computer Science& \\
Sam Gordon &Columbia& \\
Michael Grossberg &CCNY/Computer Science& \\
\hline
\end{tabular}
\end{small}
\clearpage

\begin{center} {\Large {\bf Participants cont'}} \end{center}
\begin{small}
\begin{tabular}{|p{1.4in}|p{3.0in}|p{0.5in}|}
\hline
Michael Haken &NASA/GSFC, Science Systems \& Applications, Inc.& \\
Zachary Haberman &SUNY at Albany/Physics& \\
Naveed Hasan &Columbia/Computer Science& \\
Nicholas Hatzopoulos &CUNY Gradate Center& \\
William D Heavlin &Google, Inc& speaker\\
Kay Hiranaka &CUNY Graduate Center& \\
Hao Huang &Stony Brook University/Computer Science& \\
Darryl Hutchinson &CUNY Graduate Center& \\
\hline
Ilknur Icke &CUNY Graduate Center/Computer Science& \\
Geetha Jagannathan &Columbia/Computer Science& \\
Jon Jenkins &SETI Institute/NASA Ames Research Center& speaker\\
Will Jessop &Columbia/Statistics&speaker\\
Ed Kaita &SSAI@GSFC/NASA& \\
Eyal Kazin &New York University& speaker\\
Lee Case Klippel &Columbia University& \\
Christian D. Klose &Think GeoHazards& speaker\\
Kirk Knobelspiesse &ORAU-NASA/GISS& \\
Kevin Knuth &SUNY at Albany& speaker\\
Ioannis Kostopoulos &CUNY Graduate Center& \\
\hline
Duane Lee &Columbia/Astronomy&speaker\\
Erbo Li &Columbia/Computer Science& \\
Charles Liu &CUNY at CSI& \\
Thomas Loredo &Cornell/Astronomy&speaker\\
Chris Malone &SUNY at Stony Brook/Physics and Astronomy& \\
Szabolcs Marka &Columbia/Physics& \\
Zsuzsa Marka &Columbia Astrophysics Laboratory& \\
Haley Maunu &SUNY at Albany/Physics& \\
Vinod Mohan &Columbia& \\
Claire Monteleoni &Columbia/Computer Science&speaker\\
\hline
\end{tabular}
\end{small}
\clearpage

\begin{center} {\Large {\bf Participants cont'}} \end{center}
\begin{small}
\begin{tabular}{|p{1.4in}|p{3.0in}|p{0.5in}|}
\hline
Catherine Naud &NASA/GISS& \\
Alejandro Nunez &CUNY Graduate Center& \\
Indrani Pal &IRI, The Earth Institute& \\
Vladimir Pavlovic &Rutgers/Computer Science& \\
Joshua Peek &Columbia/Astronomy& speaker\\
\hline
Fergie Ramos &Student/College of New Rochelle& \\
Khary Richardson &CUNY Graduate Center& \\
William Rossow &CCNY/NOAA-CREST& \\
Syamantak Saha &NYU Stern& \\
Destry Saul &Columbia/Astronomy& \\
Joshua Schlieder &SUNY at Stony Brook& \\
Frank Scalzo &NASA/GISS& \\
Bodhisattva Sen &Columbia/Statistics& \\
Monika Sikand &Stevens Institute of Tech/Physics& \\
Michal Simon &SUNY at Stony Brook/Physics and Astronomy& \\
Ishan Singh &Columbia& \\
Damain Sowinski &CUNY Graduate Center& \\
\hline
Ilya Tsinis &NYU/Courant& \\
Maria Tzortziou &GSFC/NASA - University of Maryland& \\
Michael Way &NASA/GISS&  \\
Adrian Weller &Columbia/Computer Science& \\
Laird Whitehill &Cornell/Astronomy& \\
Alexander Wong &JP Morgan Chase& \\
Nettie Wong &American Museum of Natural History& \\
Emmi Yonekura &Columbia University& \\
Shinjae Yoo &Brookhaven National Lab/Computer Science Center& \\
Sejong Yoon &Rutgers/Computer Science& \\
Dantong Yu &Brookhaven National Lab& \\
\hline
\end{tabular}
\end{small}

\newpage
\setcounter{section}{0}
\setcounter{figure}{0}

\begin{center}
{\Large {\bf Video Links}}
\end{center}
\bigskip

\begin{footnotesize}
\noindent
\begin{tabular}{|p{3.9in}|p{1.3in}|}
\hline
Introduction & Michael Way\\
http://www.youtube.com/watch?v=CHEoya\_e2Do&(NASA/GISS)\\
\hline
How long will it take. A historical approach to boundary crossing & Victor de la Pe\~{n}a\\
http://www.youtube.com/watch?v=3gfHeerVqHs & (Columbia U)\\
\hline
Cosmology through the Large-Scale Structure of the Universe & Eyal Kazin\\
http://www.youtube.com/watch?v=es4dH0jBJYw & (NYU)\\
\hline
On the shoulders of Gauss, Bessel, and Poisson: links, chunks, spheres, and conditional models
& William Heavlin\\
http://www.youtube.com/watch?v=FzFAxaXTkUQ & (Google Inc.) \\
\hline
Mining Citizen Science Data: Machine Learning Challenges & Kirk Borne\\
http://www.youtube.com/watch?v=XoS\_4axsb5A&(GMU)\\
\hline
Tracking Climate Models: Advances in Climate Informatics & Claire Monteleoni\\
http://www.youtube.com/watch?v=78IeffwV6bU & (Columbia U)\\
\hline
Spectral Analysis Methods for Complex Source Mixtures & Kevin Knuth\\
http://www.youtube.com/watch?v=t9mCoff5YZo & (Suny/Albany)\\
\hline
Beyond Objects: Using machines to understand the diffuse universe & Joshua Peek\\
http://www.youtube.com/watch?v=hgp8CeR43So & (Columbia U)\\
\hline
Viewpoints: A high-performance visualization and analysis tool & Michael Way\\
http://www.youtube.com/watch?v=FsW64idYw6c & (NASA/GISS)\\
\hline
%
Clustering Approach for Partitioning Directional Data in Earth and Space Sciences
& Christian Klose\\
http://www.youtube.com/watch?v=5Ak6\_rvayqg&(Think Geohazards)\\
\hline
Planetary Detection: The Kepler Mission & Jon Jenkins\\
http://www.youtube.com/watch?v=2dnA95smRE0 & (NASA/Ames)\\
\hline
Understanding the possible influence of the solar activity on the terrestrial climate:
A times series analysis approach & Elizabeth Mart\'{i}nez-G\'{o}mez\\
http://www.youtube.com/watch?v=CXmC1dh9Wdg & (Penn State)\\
\hline
Bayesian adaptive exploration applied to optimal scheduling of
exoplanet Radial Velocity observations & Tom Loredo\\
http://www.youtube.com/watch?v=B1pVtHmu9E0 & (Cornell U)\\
\hline
Bayesian Inference from Photometric Surveys & Tamas Budavari\\
http://www.youtube.com/watch?v=xCzUl0HtWGM & (JHU)\\
\hline
Long-Range Forecasts Using Data Clustering and Information Theory & Dimitris Giannakis\\
http://www.youtube.com/watch?v=pFDqP94btCg & (NYU)\\
\hline
Comparison of Information-Theoretic Methods to estimate the information flow in a dynamical
system & Deniz Gencaga\\
http://www.youtube.com/watch?v=ejX\_MWImP6A& (CCNY)\\
\hline
Reconstructing the Galactic halo's accretion history: & Duane Lee \& \\
A finite mixture model approach                       & Will Jessop \\
http://www.youtube.com/watch?v=moehYYsIOFw & (Columbia U)\\
\hline
\end{tabular}
	
\end{footnotesize}
\end{document}